\documentclass[11pt, a4paper]{article}

\usepackage{listings}
\usepackage{geometry}
\usepackage{amssymb}
\usepackage{amsmath}
\usepackage{amsthm}
\usepackage{amsfonts}
\usepackage{mathtools}
\usepackage{mathrsfs}
\usepackage{bm} 

\usepackage{graphicx}
\usepackage{dcolumn}
\usepackage[ruled,lined,algonl]{algorithm2e}
\usepackage{color}
\usepackage{autobreak}
\usepackage{subfigure}

\usepackage{CJKutf8}
\usepackage{float} 
\usepackage{authblk}

\theoremstyle{plain}
\newtheorem{theorem}{Theorem}

\newtheorem{lemma}{Lemma}
\newtheorem{proposition}{Proposition}

\theoremstyle{definition}
\newtheorem{definition}{Definition}
\newtheorem{model}{Model}

\theoremstyle{remark}
\newtheorem{remark}{Remark}

\newcommand{\field}[1]{\mathcal{#1}}
\newcommand{\pset}[1]{\mathcal{#1}}

\DeclareMathOperator{\ii}{i}

\DeclareMathOperator{\res}{res}

\DeclareMathOperator{\discr}{discr}

\newcommand{\uvar}{\bm{u}}

\begin{document}
\begin{CJK}{UTF8}{gbsn}

\title{Complex dynamics of knowledgeable monopoly models with gradient mechanisms}

\author[a]{Xiaoliang Li}

\author[b]{Jiacheng Fu}

\author[b,c]{Wei Niu\thanks{Corresponding author: wei.niu@buaa.edu.cn}}

\affil[a]{School of Digital Economics, Dongguan City University, Dongguan, China}

\affil[b]{Sino-French Engineer School, Beihang University, Beijing, China}
\affil[c]{Beihang Hangzhou Innovation Institute Yuhang, Hangzhou, China}

\date{}
\maketitle

\begin{abstract}
In this paper, we explore the dynamics of two monopoly models with knowledgeable players. The first model was initially introduced by Naimzada and Ricchiuti, while the second one is simplified from a famous monopoly introduced by Puu. We employ several tools based on symbolic computations to analyze the local stability and bifurcations of the two models. To the best of our knowledge, the complete stability conditions of the second model are obtained for the first time. We also investigate periodic solutions as well as their stability. Most importantly, we discover that the topological structure of the parameter space of the second model is much more complex than that of the first one. Specifically, in the first model, the parameter region for the stability of any periodic orbit with a fixed order constitutes a connected set. In the second model, however, the stability regions for the 3-cycle, 4-cycle, and 5-cycle orbits are disconnected sets formed by many disjoint portions. Furthermore, we find that the basins of the two stable equilibria in the second model are disconnected and also have complicated topological structures. In addition, the existence of chaos in the sense of Li-Yorke is rigorously proved by finding snapback repellers and 3-cycle orbits in the two models, respectively.

\vspace{10pt}
\end{abstract}

\section{Introduction}

Unlike a competitive market with a large number of relatively small companies producing homogeneous products and competing with each other, an oligopoly is a market supplied only by a few firms. It is well known that Cournot developed the first formal theory of oligopoly in \cite{Cournot1838R}, where players are supposed to have the naive expectations that their rivals produce the same quantity of output as in the immediately previous period.  Cournot introduced a gradient mechanism of adjusting the quantity of output and proved that his model has one unique equilibrium, which is globally stable provided that only two firms exist in the market.

A monopoly is the simplest oligopoly, which is a market served by one unique firm. In the existing literature, a market supplied by two, three, or even four companies is called a duopoly \cite{Li2022A}, a triopoly \cite{Ma2013C}, or a quadropoly \cite{Matouk2017N}, respectively. However, a monopoly may also exhibit complex dynamic behaviors such as periodic orbits and chaos if the involved firm is supposed to be boundedly rational. As distinguished by Matsumoto and Szidarovszky \cite{Matsumoto2022N}, a boundedly rational monopolist is said to be \emph{knowledgeable} if it has full information regarding the inverse demand function, and \emph{limited} if it does not know the form of the inverse demand function but possesses the values of output and price only in the past two periods. Knowledgeable and limited players have been considered in several monopoly models.

For example, Puu \cite{Puu1995T} introduced a monopoly where the inverse demand function is a cubic function with an inflection point, and the marginal cost is quadratic. In this model, the monopolist is supposed to be a limited player. Puu indicated that there exist multiple (at most three) equilibria, and complex dynamics such as chaos may appear if the reactivity of the monopolist becomes sufficiently large. Moreover, Puu's model was reconsidered by Al-Hdaibat and others in \cite{AlHdaibat2015O}, where a numerical continuation method is used to compute solutions with different periods and determine their stability regions. In particular, they analytically investigated general formulae for solutions with period four.

It should be mentioned that the equilibrium multiplicity and complex dynamics of Puu’s model might depend strictly on the inverse demand function that has an inflection point. In this regard, Naimzada and Ricchiuti \cite{Naimzada2008C} introduced a simpler monopoly with a knowledgeable player, where the inverse demand function is still cubic but has no inflection points. It was discovered that complex dynamics can also arise, especially when the reaction coefficient to variation in profits is high. Askar \cite{Askar2013O} and Sarafopoulos \cite{Sarafopoulos2015C} generalized the inverse demand function of Naimzada and Ricchiuti to a function of a similar form, but the degree of their function could be any positive integer. The difference is that the cost function in Askar's model is linear but quadratic in Sarafopoulos's.

Cavalli and Naimzada \cite{Cavalli2015E} studied a monopoly model characterized by a constant elasticity demand function, in which the firm is also assumed to be knowledgeable with a linear cost. They focused on the equilibrium stability as the variation of the price elasticity of demand and proved that there are two possible different cases, where elasticity has either a stabilizing or a mixed stabilizing/destabilizing effect. Moreover, Elsadany and Awad \cite{Elsadany2016D} explored a monopoly game with delays where the inverse demand is a log-concave function. Caravaggio and Sodini \cite{Caravaggio2020M} considered a nonlinear model, where a knowledgeable monopolist provides a fixed amount of an intermediate good and then uses this good to produce two vertically differentiated final commodities. They found that there are chaotic and multiple attractors. Furthermore, continuous dynamical systems have also been applied in the study of monopolistic markets. In \cite{Matsumoto2012N}, Matsumoto and Szidarovszky proposed a monopoly model formulated in continuous time and investigated the effect of delays in obtaining and implementing the output information. Motivated by the aforementioned work, other remarkable contributions including \cite{Gori2016D, Guerrini2018E} were done in this strand of research.

In our study, we consider two monopoly models formulated with discrete dynamical systems, where the players are supposed to be knowledgeable. The two models are distinct mainly in their inverse demand functions. The first model uses the inverse demand of Naimzada and Ricchiuti \cite{Naimzada2008C}, while the second one employs that of Puu \cite{Puu1995T}. For both models, we analyze the existence and local stability of equilibria and periodic solutions by using tools based on symbolic computations such as the method of triangular decomposition and the method of partial cylindrical algebraic decomposition. It should be mentioned that different from numerical computations, symbolic computations are exact, thus the results can be used to rigorously prove economic theorems in some sense.

The main contributions of this paper are as follows. To the best of our knowledge, the complete stability conditions of the second model are obtained for the first time. We also investigate the periodic solutions in the two models as well as their stability. Most importantly, we find different topological structures of the parameter spaces of the two considered models. Specifically, in the first model, the parameter region for the stability of any periodic solution with a fixed order constitutes a connected set. In the second model, however, the stability regions for the 3-cycle, 4-cycle, and 5-cycle orbits are disconnected sets formed by many disjoint portions. In other words, the topological structures of the regions for stable periodic orbits in Model 2 are much more complex than those in Model 1. This may be because the inverse demand function of Model 2 has an inflection point. Furthermore, according to our numerical simulations of Model 2, it is discovered that the basins of the two stable equilibria are disconnected and also have complex topological structures. In addition, the existence of chaos in the sense of Li-Yorke is rigorously proved by finding snapback repellers and 3-cycle orbits in the two models, respectively. 

The rest of this paper is organized as follows. In Section 2, we revisit the construction of the two models. In Section 3, the local stability of the equilibrium is thoroughly studied, and bifurcations through which the equilibrium loses its stability are also investigated. In Section 4, the existence and stability of periodic orbits with relatively lower orders are explored for the two models. In Section 5, we rigorously derive the existence of chaotic dynamics in the sense of Li-Yorke. The paper is concluded with some remarks in Section 6.

\section{Basic Models}

Suppose a monopolist exists in the market, and the quantity of its output is denoted as $x$. We use $P(x)$ to denote the price function (also called inverse demand function), which is assumed to be downward sloping, i.e.,
	\begin{equation}\label{eq:down}
		\frac{dP(x)}{dx}<0,~~~\text{for any}~x> 0.
	\end{equation}
It follows that $P(x)$ is invertible. The demand function (the inverse of $P(x)$) exists and is also downward sloping.  Furthermore, the cost function is denoted as $C(x)$. Then the profit is
$$\Pi(x)=P(x)x-C(x).$$

The monopolist is assumed to adopt a gradient mechanism of adjusting its output to achieve increased profits. Suppose that the firm is a knowledgeable player, which means that it has full information regarding the inverse demand function $P(x)$ and has the capability of computing the marginal profit ${{\rm d}\Pi}/{{\rm d}x}$. The firm adjusts its output by focusing on how the variation of $x$ affects the variation of $\Pi(x)$. Specifically, the adjustment process is formulated as
\begin{equation*}
	x(t+1)=x(t)+K\frac{{\rm d}\Pi(x(t))}{{\rm d}x(t)},~~~K>0.
\end{equation*}
Since $K>0$, a positive marginal profit induces the monopolist to adjust the quantity of its output in a positive direction and vice versa.

The first model considered in this paper was initially proposed by Naimzada and Ricchiuti \cite{Naimzada2008C}, where a cubic price function without the inflection point is employed. We restate the formulation of this model in the sequel.

\begin{model}
	The price function is cubic and the cost function is linear as follows.
	$$P(x)=a-bx^3,~~~C(x)=cx,$$
	where $a,b,c$ are parameters. The downward sloping condition \eqref{eq:down} is guaranteed if ${\rm d}P/{\rm d}x=-3bx^2<0$, that is if $b>0$. Moreover, assume that the marginal cost ${\rm d}C/{\rm d}x=c>0$. We adopt the general principle of setting price above marginal cost, i.e., $P(x)-c>0$ for any $x\geq 0$. Therefore, we must have that $a>c$. One knows the profit function is
	$$\Pi(x)=P(x)x-C(x)=(a-bx^3)x-cx=(a-c)x-bx^4.$$
	Thus, the gradient adjustment mechanism can be described as
	$$x(t+1)=x(t)+K(a-c-4bx^3(t)),~~~K>0.$$
	Without loss of generality, we denote $f=4bK$ and $e=(a-c)/4b$. Then, the model is simplified into a map with only two parameters:
	\begin{equation}\label{eq:m1-ef}
		x(t+1)=x(t)+f(e-x^3(t)),~~~e,f>0.
	\end{equation}
\end{model}


The second model considered in this paper is simplified from a famous monopoly model introduced by Puu \cite{Puu1995T}. We retain the same inverse demand function and cost function. The only difference is that the monopolist in our model is knowledgeable, whereas the monopolist in Puu's original model is limited.

\begin{model}
	The price function is cubic of a more general form
	$$P(x)=a_1-b_1x+c_1x^2-d_1x^3,$$
	where $a_1,b_1,c_1,d_1>0$ are parameters. The cost function is also cubic and has no fixed costs, i.e.,
	$$C(x)=a_2x-b_2x^2+c_2x^3,$$
	where $a_2,b_2,c_2>0$.
	Hence, the profit function becomes 
	$$\Pi(x)=P(x)x-C(x)=(a_1-a_2)x-(b_1-b_2)x^2+(c_1-c_2)x^3-d_1x^4,$$
	which can be denoted as
	$$\Pi(x)=ax-bx^2+cx^3-dx^4$$
	with 
	$$a=a_1-a_2, ~b=b_1-b_2, ~c=c_1-c_2,~\text{and}~d=d_1.$$
	For the sake of simplicity, we assume that $a,b,c,d>0$.
	The marginal profit $d\Pi/dx$ is directly obtained and the gradient adjustment mechanism can be formulated as
	\begin{equation}\label{eq:m2-iter}
		x(t+1)=x(t)+K(a-2bx(t)+3cx^2(t)-4dx^3(t)),~~~a,b,c,d>0.
	\end{equation}

\end{model}


\section{Local Stability and Bifurcations}

Firstly, we explain the main idea of the symbolic approach used in this paper by analyzing stepwise the local stability of Model 1. Then the theoretical results of Model 2 are reported without giving all the calculation details. 

\subsection{Model 1}

\begin{proposition}\label{prop:m1-stab}
Model 1 always has a unique equilibrium, which is stable if
$$4b(a-c)^2K^3<\frac{8}{27}$$
Moreover, there is a period-doubling bifurcation if 
$$4b(a-c)^2K^3=\frac{8}{27}.$$
\end{proposition}

The above proposition is a known result, which was first derived by Naimzada and Ricchiuti \cite{Naimzada2008C}. Indeed, this proposition can be easily proved since the analytical expression of the unique equilibrium can be obtained, i.e., $x^*=(\frac{a-c}{4b})^{1/3}.$ However, we would like to provide another proof in a computational style to demonstrate in detail how our symbolic approach works.

In what follows, the model formulation \eqref{eq:m1-ef} is taken. By setting $x(t+1)=x(t)=x$, we acquire the equilibrium equation $x= x+f(e- x^3)$. An equilibrium $x$ of the one-dimensional iteration map is locally stable if
$$\left|\frac{{\rm d}x(t+1)}{{\rm d}x(t)}\bigg|_{x(t)=x}\right|=\left|1-3fx^2\right|<1.$$
Moreover, we say the equilibrium $x$ to be feasible if $x>0$. Thus, a stable and feasible equilibrium can be characterized as a real solution of
\begin{equation}\label{eq:m1-semi}
	\left\{\begin{split}
		& x=x + f(e-x^3),\\
		& \left|1-3fx^2\right|<1,\\
		& x>0, ~e>0, ~f>0.
	\end{split}\right.
\end{equation}

Although system \eqref{eq:m1-semi} is so simple that one can solve the closed-form expression of $x$ from the equality part, the problem is how we handle a general polynomial that may have no closed-form solutions. Furthermore, it is also a nontrivial task to identify the conditions on the parameters whether a system with inequalities has real solutions. In \cite{Li2014C}, the first author of this paper and his coworker proposed an algebraic approach to systematically tackle these problems. The main idea of this approach is as follows.

The parametric system \eqref{eq:m1-semi} is univariate in $x$. For a univariate system, we introduce a key concept called \emph{border polynomial} in the sequel. One useful property of a border polynomial is that its real zeros divide the parameter space into separated regions and the solution number of the original system is invariant for all parameter points in each region.

\begin{definition}[Border Polynomial]\label{def:bp}
Consider a univariate system
\begin{equation}\label{ex:uni-pq}
\left\{
\begin{array}{l}\smallskip
P(\uvar,x)=\sum_{i=0}^m a_i(\uvar)\,x^i=0,\\
Q_1(\uvar,x)>0,\ldots,Q_s(\uvar,x)>0,
\end{array}
\right.
\end{equation}
where $P$ and $Q_1,\ldots,Q_s$ are univariate polynomials in $x$, and $\uvar$ stands for all parameters. 
The product
$$a_m(\uvar)\cdot\discr(P)\cdot\prod_{i=1}^s\res(P,Q_i)$$ is called
the \emph{border polynomial} of system \eqref{ex:uni-pq}. Here, $\res(F,G)$ stands for the resultant of two polynomials $F$ and $G$, while $\discr(F)$ denotes the discriminant of $F$.
\end{definition}

More specifically, the formal definitions of the resultant and the discriminant in the above definition are given as follows. Let $$F=\sum_{i=0}^ma_i\,x^i,\quad G=\sum_{j=0}^lb_j\,x^j$$ be two univariate polynomials in $x$ with coefficients $a_i,b_j$ in the field of complex numbers, and $a_m,b_l\neq 0$. The
determinant
\begin{equation*}\label{eq:sylmat}
 \begin{array}{c@{\hspace{-5pt}}l}
 \left|\begin{array}{cccccc}
a_m & a_{m-1}& \cdots   & a_0   &        &       \\
           & \ddots   & \ddots&    \ddots    &\ddots&   \\
         &          & a_m   & a_{m-1}&\cdots& a_0 \\ [5pt]
b_l & b_{l-1}& \cdots   &  b_0 &    &         \\
            & \ddots   &\ddots &   \ddots     &\ddots&       \\
         &   &    b_{l}     & b_{l-1} & \cdots &  b_0
\end{array}\right|
& \begin{array}{l}\left.\rule{0mm}{8mm}\right\}l\\
\\\left.\rule{0mm}{8mm}\right\}m
\end{array}
\end{array}
\end{equation*}
is called the \emph{Sylvester resultant} (or simply
\emph{resultant}) of $F$ and $G$, and denoted by $\res(F,G)$. 
The resultant of $F$ and its derivative ${\rm d}F/{\rm d}x$, i.e., $\res(F,{\rm d}F/{\rm d}x)$, is called the \emph{discriminant} of $F$ and denoted by $\discr(F)$. The following lemma is one of the well-known properties of resultants, which could be found in \cite{Mishra1993A}.

\begin{lemma}\label{lem:res-com}
 Two univariate polynomials $F$ and $G$ have common zeros in the field of complex numbers if and only if $\res(F,G)=0$. Moreover, a univariate polynomial $F$ has a multiple zero in the field of complex numbers if and only if $\discr(F)=0$.
\end{lemma}

It is worth noticing that the number of real zeros of $P$ may change when the leading coefficient $a_m(\uvar)$ or the discriminant $\discr(P)$ goes from non-zero to zero and vice versa. In addition, if $\res(P, Q_i)$ goes across zero, then the zeros of $P$ will pass through the boundaries of $Q_i>0$, which means that the number of real roots of \eqref{ex:uni-pq} may change. Therefore, the following lemma is derived.

\begin{lemma}\label{lem:main}
Consider a univariate system as \eqref{ex:uni-pq}. Let $A$ and $B$ be two points in the space of parameters $\uvar$. Suppose that any of $A$, $B$ does not annihilate the border polynomial of system \eqref{ex:uni-pq}. If there exists a real path $C$ from $A$ to $B$ such that any point on $C$ is not a root of the border polynomial, then the number of real solutions of system \eqref{ex:uni-pq} evaluated at $A$ is the same as that at $B$. 
\end{lemma}

\begin{figure}[htbp]
    \centering
    \includegraphics[width=10cm]{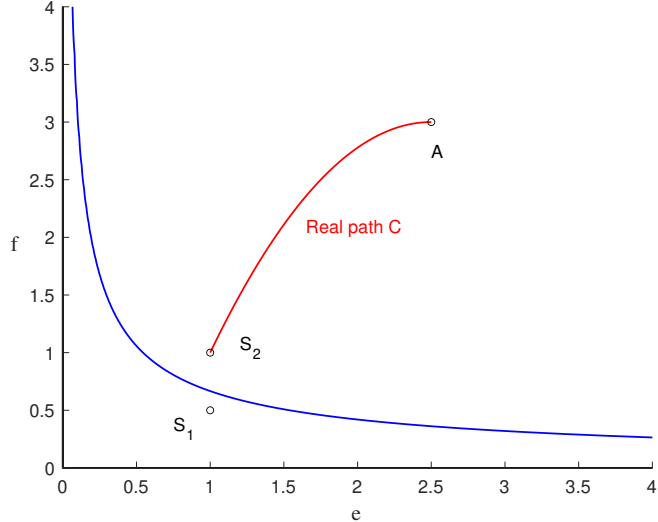}
    \caption{Partitions of the parameter space of Model 1 and sample points}
    \label{fig:par-space-model2}
\end{figure}

Since $1-3fx^2<1$, we know that system \eqref{eq:m1-semi} is equivalent to 
\begin{equation}\label{eq:m1-semi-simple}
	\left\{\begin{split}
		& x^3-e=0,\\
		& 2-3fx^2>0,\\
		& x>0, ~e>0, ~f>0.
	\end{split}\right.
\end{equation}
We have $a_m=1$ and $\discr(x^3-e)=27e^2$. Moreover, $\res(x^3-e, 2-3fx^2)= -27e^2f^3 + 8$ and $\res(x^3-e, x)=e$. According to Definition \ref{def:bp}, the border polynomial of system \eqref{eq:m1-semi-simple} is $27e^3(-27e^2f^3+8)$, the zeros of which are marked in blue as shown in Figure \ref{fig:par-space-model2}. This blue curve divides the parameter set $\{(e,f)\,|\,e>0,f>0\}$ into two (the northeast and the southwest) regions. 

Notice the two points $S_2$ and $A$ in Figure \ref{fig:par-space-model2}. One can find a real path $C$ from $A$ to $S_2$ such that it does not pass through the blue curve. According to Lemma \ref{lem:main},  system \eqref{eq:m1-semi-simple} has the same number of real roots with the parameters evaluated at $S_2$ and $A$. This means that the number of real solutions of system \eqref{eq:m1-semi-simple} is invariant in the northeast region. Therefore, we can choose a sample point from each region to determine the root number. For this simple system, sample points might be selected directly by eyes, e.g., $S_1=(1, 1/2)$, $S_2=(1,1)$. However, the choosing process might be extremely complex in general, which could be done automatically by using, e.g., the method of partial cylindrical algebraic decomposition or called the PCAD method \cite{Collins1991P}.

For each region, one can determine the root number by counting roots of the non-parametric system of \eqref{eq:m1-semi-simple} evaluated at the corresponding sample point. Take $S_1$ as an example, where \eqref{eq:m1-semi-simple} becomes
\begin{equation}\label{eq:no-par-sys}
	\left\{x^3-1=0,~2-\frac{3}{2}x^2>0,~x>0\right\}.
\end{equation}
In order to count the number of its real roots, an obvious way is directly solving $x^3-1=0$, i.e., $x=1$, and then checking whether $2-\frac{3}{2}x^2>0$ and $x>0$ are satisfied. The result is true, which means that there exists one unique real solution of \eqref{eq:no-par-sys}. However, it is difficult to precisely obtain all real zeros of a general univariate system since root formulae do not exist for polynomials with degrees greater than $4$. Therefore, a more systematic method called real root counting \cite{Xia2002A} is generally needed here, and we demonstrate how this method works by using \eqref{eq:no-par-sys} as an example.

It is noted that $x^3-1$, $2-\frac{3}{2}x^2$ and $x$ have no common zeros, i.e., they have no factors in common. Otherwise, one needs to reduce the common factors from the inequalities first. After that, we isolate all real zeros of $2-\frac{3}{2}x^2$ and $x$ by rational intervals, e.g., 
\begin{equation}\label{eq:close-int}
    \left[-\frac{12}{10},-\frac{11}{10}\right], ~\left[-\frac{1}{10},\frac{1}{10}\right], ~\left[\frac{11}{10}, \frac{12}{10}\right].
\end{equation}
Although it is trivial for this simple example, the isolation process could be particularly tough for general polynomials, which may be handled by using, e.g., the modified Uspensky algorithm \cite{Collins1983R}. Moreover, the intervals can be made as small as possible to guarantee no zeros of $x^3-1$ lie in these intervals, which could be checked by using, e.g., Sturm's theorem \cite{Sturmfels2002S}. Thus, the real zeros of $x^3-1$ must be in the complement of \eqref{eq:close-int}:
\begin{equation}\label{eq:open-interval}
	\left(-\infty,-\frac{12}{10}\right),~\left(-\frac{11}{10},-\frac{1}{10}\right),~\left(\frac{1}{10},\frac{11}{10}\right),~\left(\frac{12}{10},+\infty\right).
\end{equation}

In each of these open intervals, the signs of $2-\frac{3}{2}x^2$ and $x$ are invariant and can be determined by checking them at selected sample points. For instance, to determine the sign of $2-\frac{3}{2}x^2$ on $(12/10,+\infty)$, we check the sign at a sample point, e.g., $x=2$. We have that $2-\frac{3}{2}x^2|_{x=2}=-4<0$, thus $2-\frac{3}{2}x^2<0$ on $(12/10,+\infty)$. Similarly, it is obtained that the signs of $2-\frac{3}{2}x^2$ and $x$ at \eqref{eq:open-interval} are $-,+,+,-$ and $-,-,+,+$, respectively. Hence, $(1/10,11/10)$ is the only interval such that the two inequalities $2-\frac{3}{2}x^2>0$ and $x>0$ of system \eqref{eq:no-par-sys} are simultaneously satisfied.

We focus on $(1/10,11/10)$. Using Sturm's theorem,  we can count the number of the real zeros of $x^3-1$ at $(1/10,11/10)$, which is one.  Therefore, system \eqref{eq:m1-semi-simple} has one real root at $S_1=(1, 1/2)$. The above approach works well for a system formulated with univariate polynomial equations and inequalities although some steps seem silly and not necessary for this simple example. Similarly, we know that system \eqref{eq:m1-semi-simple} has no real roots at $S_2=(1, 1)$. 

In conclusion, system \eqref{eq:m1-semi-simple} has one real root if the parameters take values from the southwest region where $S_1$ lies, and has no real roots if the parameters take values from the northeast region where $S_2$ lies. Furthermore, the inequalities of some factors of the border polynomial may be used to explicitly describe a given region. It is evident that $27e^2f^3-8<0$ describes the region where $S_1$ lies. Therefore, Model 1 has one unique stable equilibrium provided that
 $$e^2f^3=\left(\frac{a-c}{4b}\right)^2(4bK)^3=4b(a-c)^2K^3<\frac{8}{27},$$
which is consistent with Proposition \ref{prop:m1-stab}.

According to the classical bifurcation theory, for a one-dimensional iteration map $x(t+1)=F(x(t))$, we know that bifurcations may occur if 
$$\left|\frac{{\rm d}x(t+1)}{{\rm d}x(t)}\bigg|_{x(t)=x}\right|=\left|\frac{{\rm d}F}{{\rm d}x}\right|=1.$$
More specifically, if ${\rm d}F/{\rm d}x=-1$, then the system may undergo a period-doubling bifurcation (also called flip bifurcation), where the dynamics switch to a new behavior with twice the period of the original system. On the other hand, if ${\rm d}F/{\rm d}x=1$, then the system may undergo a saddle-node (fold), transcritical, or pitchfork bifurcation. One might determine the type of bifurcation from the change in the number of the (stable) equilibria. In the case of saddle-node bifurcation, one stable equilibrium (a node) annihilates with another unstable one (a saddle). Before and after a transcritical bifurcation, there is one unstable and one stable equilibrium, and the unstable equilibrium becomes stable and vice versa. In the case of pitchfork bifurcation, the number of equilibria changes from one to three or from three to one, while the number of stable equilibria changes from one to two or from one to zero. Accordingly, it is concluded that Model 1 may undergo a period-doubling bifurcation if 
$$e^2f^3=4b(a-c)^2K^3=\frac{8}{27},$$
and there are no other bifurcations.

\subsection{Model 2}

According to \eqref{eq:m2-iter}, by setting $x(t+1)=x(t)=x$, we know that Model 2 has at most three equilibria. The analytical expressions of the equilibria exist, but are complex, i.e.,
\begin{equation}\label{eq:closed-e3}
    \begin{split}
x_1=\, & {\frac {\sqrt [3]{M}}{12d}} -{\frac {8bd-3{c}^{2}}{4d\sqrt[3]{M}}}+{\frac {c}{4d}},\\
x_{2,3}=\, & -{\frac {\sqrt [3]{M}}{24d}}+{\frac {8bd-3{c}^{2}}{8d\sqrt [3]{M}}}+{\frac {c}{4d}}\pm \frac{\ii\sqrt{3}}{2} \left( {\frac {\sqrt [3]{M}}{12d}}+{\frac {8bd-3{c}^{2}}{4d\sqrt [3]{M}}} \right),
    \end{split}
\end{equation}
where
\begin{align*}
M=12d \sqrt{3} \sqrt{108{a}^{2}{d}^{2}-108abcd+27\,a{c}^{3}+32{b}^{3}d-9{b}^{2}{c}^{2}}+216a{d}^{2}-108bcd+27{c}^{3}.
\end{align*}

Furthermore, an equilibrium $x$ is locally stable provided that
$$\left|\frac{{\rm d}x(t+1)}{{\rm d}x(t)}\bigg|_{x(t)=x}\right|=
	\left|1+K(-2b+6cx-12dx^2)\right|<1.$$
Hence, a stable equilibrium of map \eqref{eq:m2-iter} is a real solution of 
\begin{equation}\label{eq:semi-mod2}
	\left\{\begin{split}
		& x=x + K(a-2bx+3cx^2-4dx^3),\\
		& K(-2b+6cx-12dx^2)<0,\\
		& 2+K(-2b+6cx-12dx^2)>0,\\
		& x>0, ~a>0,~ b>0,~ c>0,~ d>0.
	\end{split}\right.
\end{equation}

Obviously, analyzing the stable equilibrium by substituting the closed-form solutions \eqref{eq:closed-e3} into \eqref{eq:semi-mod2} is complicated and impractical. In comparison, the approach applied in the analysis of Model 1 does not require explicitly solving any closed-form equilibrium. If the analytical solution has a complicated expression or even if there are no closed-form solutions, our approach still works in theory.

Concerning the border polynomial of system \eqref{eq:semi-mod2}, we compute
\begin{align*}
&\discr(K(a-2bx+3cx^2-4dx^3))=-16 K^{5} d R_1,\\
&\res(K(a-2bx+3cx^2-4dx^3),K(-2b+6cx-12dx^2))= -16 K^5 d R_1,\\
&\res(K(a-2bx+3cx^2-4dx^3),2+K(-2b+6cx-12dx^2))= -16 K^2 d R_2 ,\\
&\res(K(a-2bx+3cx^2-4dx^3),x)= -K a,
\end{align*}
where 
\begin{align*}
	R_1=\,&108a^2d^2-108abcd+27ac^3+32b^3d-9b^2c^2,\\
	R_2=\,&108K^3a^2d^2-108K^3abcd+27K^3ac^3+32K^3b^3d-9K^3b^2c^2-24Kbd+9Kc^2-8d.
\end{align*}
Therefore, the border polynomial is $-16384\, d^{4} K^{14} a R_1^2 R_2$,
the zeros of which divide the parameter set $\{(a,b,c,d,K)\,|\,a,b,c,d,K>0\}$ into separated regions. The PCAD method \cite{Collins1991P} permits us to select at least one sample point from each region. In Table \ref{tab:sample-mod2}, we list the 30 selected sample points and the corresponding numbers of distinct real solutions of system \eqref{eq:semi-mod2}.

\begin{table}[htbp]
	\centering 
	\caption{Selected Sample Points in $\{(a,b,c,d,K)\,|\,a,b,c,d,K>0\}$}
	\label{tab:sample-mod2} 
	\begin{tabular}{|l|c|c|c||l|c|c|c|}  
\hline  
$(a,b,c,d,K)$ & num & $R_1$ & $R_2$ & $(a,b,c,d,K)$ & num & $R_1$ & $R_2$ \\ \hline
$(1, 1, 1/4, 1/64, 1/2)$& 2 & $-$ & $-$ & $(1, 1, 1/4, 1/64, 1)$& 1 & $-$ & $+$ \\ \hline
$(1, 1, 1/4, 1/64, 2)$& 0 & $-$ & $-$ & $(1, 1, 1/4, 19/1024, 1)$& 2 & $-$ & $-$ \\ \hline
$(1, 1, 1/4, 19/1024, 2)$& 1 & $-$ & $+$ & $(1, 1, 1/4, 19/1024, 3)$& 0 & $-$ & $-$ \\ \hline
$(1, 1, 1/4, 1/16, 1)$& 1 & $+$ & $-$ & $(1, 1, 1/4, 1/16, 2)$& 0 & $+$ & $+$ \\ \hline
$(1, 1, 1/4, 1, 1/2)$& 1 & $+$ & $-$ & $(1, 1, 1/4, 1, 1)$& 0 & $+$ & $+$ \\ \hline

$(1, 1, 3/8, 1/64, 1/8)$& 1 & $+$ & $-$ & $(1, 1, 3/8, 1/64, 1)$& 0 & $+$ & $+$ \\ \hline
$(1, 1, 3/8, 1/32, 1/4)$& 2 & $-$ & $-$ & $(1, 1, 3/8, 1/32, 1)$& 1 & $-$ & $+$ \\ \hline
$(1, 1, 3/8, 1/32, 17)$& 0 & $-$ & $-$ & $(1, 1, 3/8, 49/1024, 1)$& 2 & $-$ & $-$ \\ \hline
$(1, 1, 3/8, 49/1024, 4)$& 1 & $-$ & $+$ & $(1, 1, 3/8, 49/1024, 8)$& 0 & $-$ & $-$ \\ \hline
$(1, 1, 3/8, 1/16, 1)$& 1 & $-$ & $+$ & $(1, 1, 3/8, 1/16, 3)$& 0 & $-$ & $-$ \\ \hline

$(1, 1, 3/8, 1, 1/2)$& 1 & $+$ & $-$ & $(1, 1, 3/8, 1, 1)$& 0 & $+$ & $+$ \\ \hline
$(1, 1, 15/32, 1/16, 1/2)$& 1 & $+$ & $-$ & $(1, 1, 15/32, 1/16, 1)$& 0 & $+$ & $+$ \\ \hline
$(1, 1, 15/32, 3/32, 1)$& 1 & $+$ & $-$ & $(1, 1, 15/32, 3/32, 8)$& 0 & $+$ & $+$ \\ \hline
$(1, 1, 15/32, 1, 1/2)$& 1 & $+$ & $-$ & $(1, 1, 15/32, 1, 1)$& 0 & $+$ & $+$ \\ \hline
$(1, 1, 1, 1, 1/2)$& 1 & $+$ & $-$ & $(1, 1, 1, 1, 1)$& 0 & $+$ & $+$ \\ \hline
	\end{tabular}
\end{table}

According to Table \ref{tab:sample-mod2}, one can see that system \eqref{eq:semi-mod2} has one real solution if and only if $R_1<0, R_2>0$ or $R_1>0, R_2<0$. Moreover, a necessary condition that system \eqref{eq:semi-mod2} has two real solutions is that $R_1<0$ and $R_2<0$, which is not a sufficient condition, however. For example, at $(a,b,c,d,K)=(1, 1, 1/4, 1/64, 2)$, system \eqref{eq:semi-mod2} has no real solutions but $R_1<0$ and $R_2<0$ are fulfilled. To acquire the necessary and sufficient condition, additional polynomials ($R_3$ and $R_4$) are needed, which can be found in the so-called generalized discriminant list and can be picked out by repeated trials. Regarding the generalized discriminant list, readers may refer to \cite{Yang2001A} for more details. Due to space limitations, we directly report below the necessary and sufficient condition that system \eqref{eq:semi-mod2} has two real solutions without giving the calculation details:
$$R_1 < 0, R_2 < 0, R_3 > 0, R_4 < 0,$$
where
\begin{align*}
	R_3=\,&8Kbd-3Kc^2+8d,\\
	R_4 =\,&432K^2a^2d^3-432K^2abcd^2+108K^2ac^3d
	+128K^2b^3d^t2-36K^2b^2c^2d+192Kb^2d^2\\
	&-144Kbc^2d+27Kc^4+64bd^2-24c^2d.
\end{align*}

We continue to analyze the bifurcations of this model. An equilibrium $x$ of map \eqref{eq:m2-iter} may undergo a period-doubling bifurcation if 
$$\frac{{\rm d}x(t+1)}{{\rm d}x(t)}\bigg|_{x(t)=x}=
	1+K(-2b+6cx-12dx^2)=-1.$$
Hence, a period-doubling bifurcation may occur if the following system has at least one real solution.
\begin{equation}\label{eq:semi-mod2bf}
	\left\{\begin{split}
		& x=x + K(a-2bx+3cx^2-4dx^3),\\
		& K(-2b+6cx-12dx^2)+2=0,\\
		& x>0, ~a>0,~ b>0,~ c>0,~ d>0.
	\end{split}\right.
\end{equation}
By using the method of triangular decomposition\footnote{The method of triangular decomposition can be viewed as an extension of the method of Gaussian elimination. The main idea of both methods is to transform a system into a triangular form. However, the triangular decomposition method is available for polynomial systems, while the Gaussian elimination method is just for linear systems. Refer to \cite{Wu1986B, Li2010D, Jin2013A, Wang2001E} for more details.}, we transform the solutions of the first two equations of system \eqref{eq:semi-mod2bf} into zeros of the triangular set
\[\pset{T}=[(8Kbd-3Kc^2+4d)x-6adK+bcK-c, R_2].\]
 Obviously, the system $\{\pset{T}=0,~ x>0, ~a>0,~ b>0,~ c>0,~ d>0\}$ has at least one real positive solution if $R_2=0$ and $x=(6adK-bcK+c)/(8Kbd-3Kc^2+4d)>0$, i.e.,
\[R_2=0, ~R_5>0,\]
 where
\begin{align*}
R_5&=\,(6adK-bcK+c)(8Kbd-3Kc^2+4d)\\
&=\,48K^2abd^2-18K^2ac^2d-8K^2b^2cd+3K^2bc^3+24Kad^2+4Kbcd-3Kc^3+4cd.
\end{align*}

Similarly, concerning the occurrence of a pitchfork bifurcation, we consider 
\begin{equation}\label{eq:semi-mod2bf2}
	\left\{\begin{split}
		& x=x + K(a-2bx+3cx^2-4dx^3),\\
		& K(-2b+6cx-12dx^2)=0,\\
		& x>0, ~a>0,~ b>0,~ c>0,~ d>0,
	\end{split}\right.
\end{equation}
and count the number of stable equilibria. More details are not reported here due to space limitations. We summarize all the obtained results in the following theorem.

\begin{theorem}\label{thm:mod2-stab}
	Model 2 has at most two stable equilibria. Specifically, there exists just one stable equilibrium if 
	$$R_1<0, R_2>0~\text{or}~R_1>0, R_2<0,$$ and there exist two stable equilibria if 
	$$R_1 < 0, R_2 < 0, R_3 > 0, R_4 < 0.$$
Moreover, there is a period-doubling bifurcation if 
\[R_2=0, ~R_5>0,\]
and there is a pitchfork bifurcation if 
\[R_1=0,~ R_2>0,~R_6>0~\text{or}~R_1=0,~ R_2>0,~ R_4<0,~R_6>0,\]
where 
\begin{align*}
R_6=\,&48abd^2-18ac^2d-8b^2cd+3bc^3.
\end{align*}
\end{theorem}

\begin{figure}[htbp]
    \centering
    \includegraphics[width=10cm]{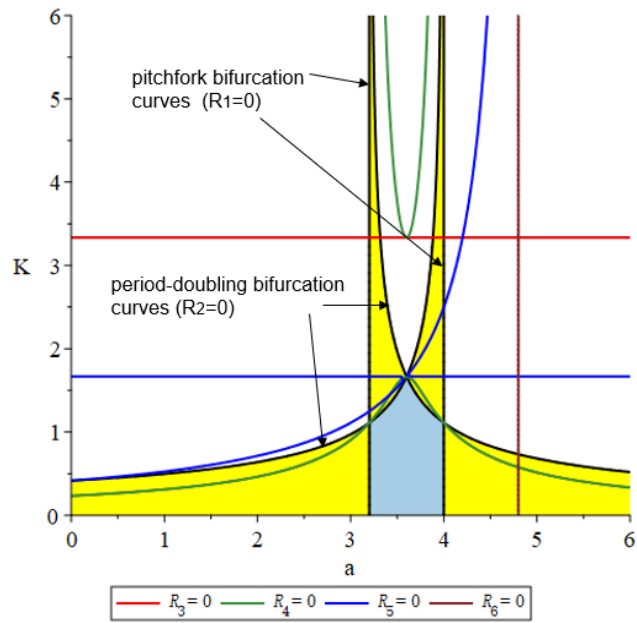}
    \caption{The two-dimensional $(a,K)$ parameter plane of Model 2 with the other parameters fixed: $b=2.4$, $c=0.6$, and $d=0.05$. The region for the existence of one stable equilibrium is colored in yellow, while that of two stable equilibria is colored in blue-gray.}
    \label{fig:par-plane-mod2}
\end{figure}

\remark{To the best of our knowledge, the stability results regarding the parameters $a,b,c,d,K$ reported in Theorem \ref{thm:mod2-stab} are new although the special case of $a=3.6$, $b=2.4$, $c=0.6$, $d=0.05$ has been discussed in \cite{Matsumoto2022N}. The two parameters $K,a$ play more ambitious roles than others in practice for $K$ controls the speed of adjusting the monopolist's output and $a$ is the difference between the initial product price of the market without any supply and the initial marginal cost of the firm without any production. By fixing $b=2.4$, $c=0.6$ and $d=0.05$, we depict the $(a,K)$ parameter plane in Figure \ref{fig:par-plane-mod2}, where the region for the existence of one stable equilibrium is colored in yellow, while the region for the existence of two stable equilibria is colored in blue-gray. Model 2 behaves differently from typical oligopolies with gradient mechanisms. As shown by Figure \ref{fig:par-plane-mod2}, for instance, even if the adjustment speed $K$ is quite large, there always exist some values of $a$ such that Model 2 is stable. Moreover, for a fixed value of $K$ greater than around $1.7$, Model 2 undergoes from instability to stability and then back to instability twice as the parameter $a$ changes from low to high.}

\section{Periodic Solutions}

From an economic point of view, it is realistic to assume that a boundedly rational firm can not learn the pattern behind output and profits if periodic dynamics take place. In this regard, we investigate the existence and stability of periodic solutions with relatively lower orders in this section.

Let $I$ be an interval of real numbers, and let $F: I\rightarrow \field{R}$ be a function. If $x\in I$,  suppose that $F^0(x)$ represents $x$ and $F^{n+1}(x)$ denotes $F(F^n(x))$ for $n\in\{0,1,\ldots\}$. A point $p\in I$ is said to be a \emph{periodic point} with period $n$ or order $n$ if $p=F^n(p)$, and $p\neq F^k(p)$ for any $1\leq k<n$. If $p$ is a point with period $n$, we call $p\mapsto F^1(p)\mapsto\cdots\mapsto F^n(p)= p$ a \emph{$n$-cycle orbit}. Furthermore, a point $y\in I$ with period $k$ is said to be \emph{asymptotically stable} if there exists $\delta$ such that $|F^k(x)-y|<|x-y|$ for all $x\in (y-\delta, y+\delta)$. 
	
The following lemma can be found in \cite{Li1975P}, which provides an algebraic criterion to verify the stability of a periodic point.

\begin{lemma}\label{lem:stable}
	Assume that $y\in I$ is a periodic point of $F$ with period $k$. If $F$ is differentiable at the points $y, F(y),\ldots, F^{k-1}(y)$, then $y$ is asymptotically stable if 
	$$\left|\prod_{i=0}^{k-1}\frac{{\rm d}}{{\rm d}x}F(y_i)\right|<1, ~~~\text{where}~y_i=F^i(y).$$
\end{lemma}

\subsection{Model 1}

We start by considering the existence of periodic orbits with order two. Assume that there is a 2-cycle orbit $x\mapsto y\mapsto x$, where $\mapsto$ stands for the iteration map \eqref{eq:m1-ef}. Thus, we have
\begin{equation}\label{eq:epart-2cycle}
y=x + f(e-x^3),~~~ x=y + f(e-y^3).
\end{equation}
Obviously, $x\neq y$ should be guaranteed. Otherwise, $x\mapsto y\mapsto x$ will degenerate into an equilibrium. Then, the problem of determining the existence of 2-cycles is transformed into determining the existence of real solutions of 
\begin{equation}\label{eq:p2m1}
	\left\{\begin{split}
		& y=x + f(e-x^3),\\
		& x=y + f(e-y^3),\\
		& x\neq y,\\
		& x>0, ~y>0, ~e>0, ~f>0.
	\end{split}\right.
\end{equation}

Since the above system involves two variables $x$ and $y$, the approach used in Section 3 (feasible only for univariate systems) might not be directly employed herein. 

\begin{remark}\label{rm:linearization}
However, we can transform system \eqref{eq:p2m1} equivalently into univariate systems based on its triangular decomposition. Specifically, the triangular decomposition method permits us to decompose the equation part \eqref{eq:epart-2cycle} into the following two triangular sets.
\begin{align*}
	&\pset{T}_{11}=[y-x, x^3-e],\\
	&\pset{T}_{12}=[y+fx^3-x-ef, f^3x^6-3f^2x^4-2ef^3x^3+3fx^2+3ef^2x+e^2f^3-2].
\end{align*}
Since the first polynomial in $\pset{T}_{11}$ is $y-x$, which implies that $x=y$. Thus, the zeros of $\pset{T}_{11}$ are not of our concern. We only focus on $\pset{T}_{12}$, where the first polynomial $y+fx^3-x-ef$ has degree one with respect to $y$. Therefore, one can directly solve $y=-fx^3+x+ef$ and substitute it into relative inequalities of system \eqref{eq:p2m1}. In short, system \eqref{eq:p2m1} can be equivalently transformed into the following univariate system.
\begin{equation*}
	\left\{\begin{split}
		& f^3x^6-3f^2x^4-2ef^3x^3+3fx^2+3ef^2x+e^2f^3-2=0,\\
		& -fx^3+x+ef>0,\\
		& x>0, ~e>0, ~f>0.
	\end{split}\right.
\end{equation*}
 After that, the approach in Section 3 can be applied. The results show that the above system has two real solutions if and only if $8/27<e^2f^3<2$. It is evident that these two real solutions belong to the same 2-cycle orbit because $x,y$ are symmetric and can be replaced with each other. Therefore, there exists at most one 2-cycle orbit in Model 1.
\end{remark}

According to Lemma \ref{lem:stable}, to determine whether the discovered 2-cycle is stable, we consider \eqref{eq:p2m1} together with the condition
$$\left|\frac{{\rm d}(x + f(e-x^3))}{{\rm d}x}\times\frac{{\rm d}(y + f(e-y^3))}{{\rm d}y}\right|<1,$$
i.e.,
$$\left|(1-3fx^2)(1-3fy^2)\right|<1.$$
The technique introduced in Remark \ref{rm:linearization} is needed to transform the system into a univariate one. According to our calculations, the unique 2-cycle orbit is stable if and only if $729e^4f^6-3294e^2f^3+1664>0$ or equivalently $8/27<e^2f^3<{(61-11\sqrt{17})}/{27}$. We collect the aforementioned results in the following theorem.

\begin{theorem}
Model 1 has at most one 2-cycle orbit, which exists if 
$$8/27<e^2f^3<2.$$ 
Furthermore, this unique 2-cycle is stable if 
$$8/27<e^2f^3<\frac{61-11\sqrt{17}}{27},$$
or approximately
$$0.2962962963<e^2f^3<0.5794754859.$$
\end{theorem}

The measurement of the magnitude of periodic orbits is economically interesting for it characterizes the size of fluctuations in dynamic economies. For a $n$-cycle orbit $p_1\mapsto p_2\mapsto \cdots p_n\mapsto p_1$, a direct definition of the magnitude measure is 
$$d=|p_1-p_2|+|p_2-p_3|+\cdots+|p_{n-1}-p_n|+|p_n-p_1|.$$
However, to obtain better mathematical properties, we square each item and define the magnitude measure to be 
$$d=(p_1-p_2)^2+(p_2-p_3)^2+\cdots+(p_{n-1}-p_n)^2+(p_n-p_1)^2.$$
For a 2-cycle orbit $x\mapsto y\mapsto x$ in Model 1, the magnitude measure becomes $d=(x-y)^2+(y-x)^2$. Thus, we have
\begin{equation*}
    \left\{ 
    \begin{split}
    & d-(x-y)^2-(y-x)^2=0,\\
    & -y+x+f(e-x^3)=0,\\
    & -x+y+f(e-y^3)=0.
    \end{split}
    \right.
\end{equation*}
Using the method of triangular decomposition, we decompose the solutions of the above system into zeros of the following two triangular sets.
\begin{equation*}
\begin{split}
\pset{T}_{21}=[\,&y-x, x^3-e, d\,],\\ 
 \pset{T}_{22}=[\,&y+x^3f-ef-x,\\
 &(d^2f^3+4df^2+4f)x^2+(-6def^3-12ef^2)x+36e^2f^3-2f^2d^2-8fd-8,\\
 &f^3d^3-12f^2d^2-60fd+216e^2f^3-64\,].
\end{split}
\end{equation*}
The first polynomial $y-x$ in $\pset{T}_{21}$ implies that $x=y$. Thus, $\pset{T}_{21}$ is not of concern since it corresponds to equilibria rather than 2-cycle orbits. We focus on the last polynomial $f^3d^3-12f^2d^2-60fd+216e^2f^3-64$ in $\pset{T}_{22}$. By solving $d$ from this polynomial, we obtain three solutions:
$$d_1=\frac{2}{f}\left(\frac{3H}{2}+\frac{6}{H}+2\right),~~d_2, d_3=\frac{2}{f}\left(-\frac{3H}{4}-\frac{3}{H}+2\pm \frac{\ii\sqrt{3}}{2}\Big(\frac{3H}{2}-\frac{6}{H}\Big)\right),$$
where 
$$H=\sqrt[3]{8-4e^2f^3+4(e^4f^6-4e^2f^3)^{1/2}}.$$
Here, only the real solution $d_1$ is meaningful. Therefore, the magnitude measure of the unique 2-cycle orbit in Model 1 can be expressed as
$$d=\frac{2}{f}\left(\frac{3\sqrt[3]{8-4e^2f^3+4(e^4f^6-4e^2f^3)^{1/2}}}{2}+\frac{6}{\sqrt[3]{8-4e^2f^3+4(e^4f^6-4e^2f^3)^{1/2}}}+2\right).$$

In the rest of this section, similar calculations as above are repeated. We omit these computation details due to space limitations. Concerning 3-cycle orbits in Model 1, we need to count real solutions of
\begin{equation*}
	\left\{\begin{split}
		& y=x + f(e-x^3),\\
		& z=y + f(e-y^3),\\
		& x=z + f(e-z^3),\\
		&  {x\neq y,~x\neq z,}\\
		& x>0, ~y>0, ~z>0, ~e>0, ~f>0.
	\end{split}\right.
\end{equation*}
Based on a series of computations, we derive the following theorem.
\begin{theorem}
Model 1 has no 3-cycle orbits for all possible parameter values such that $e,f>0$.
\end{theorem}

For a 4-cycle orbit $x\mapsto y\mapsto z \mapsto w\mapsto x$, we have the system
\begin{equation*}
	\left\{\begin{split}
		& y=x + f(e-x^3),\\
		& z=y + f(e-y^3),\\
		& w=z + f(e-z^3),\\
		& x=w + f(e-w^3),\\
		& {x\neq y,~x\neq z,~x\neq w,}\\
		& x>0, ~y>0, ~z>0, ~e>0, ~f>0.
	\end{split}\right.
\end{equation*}
Furthermore, the following condition is required to guarantee that the considered 4-cycle is stable.
$$\left|\frac{{\rm d} (x + f(e-x^3))}{{\rm d} x}\times\frac{{\rm d}(y + f(e-y^3))}{{\rm d}y}\times\frac{{\rm d}(z + f(e-z^3))}{{\rm d}z}\times\frac{{\rm d}(w + f(e-w^3))}{{\rm d} w}\right|<1,$$
i.e.,
$$\left|(1-3fx^2)(1-3fy^2)(1-3fz^2)(1-3fw^2)\right|<1.$$
As the polynomials involved in the conditions of the existence and stability of 4-cycle orbits are extremely complicated, we report below the obtained results in an approximate style.

\begin{theorem}\label{thm:mod1-4cycle}
    Model 1 has at most one 4-cycle orbit, which exists if  
    $$0.5794754859< e^2f^3< 1.237575627.$$
    Furthermore, this unique 4-cycle is stable if 
    $$0.5794754859<e^2f^3<0.6673871142.$$
\end{theorem}

\begin{figure}[htbp]
  \centering
    \subfigure[phase diagram.]{\includegraphics[width=0.49\textwidth]{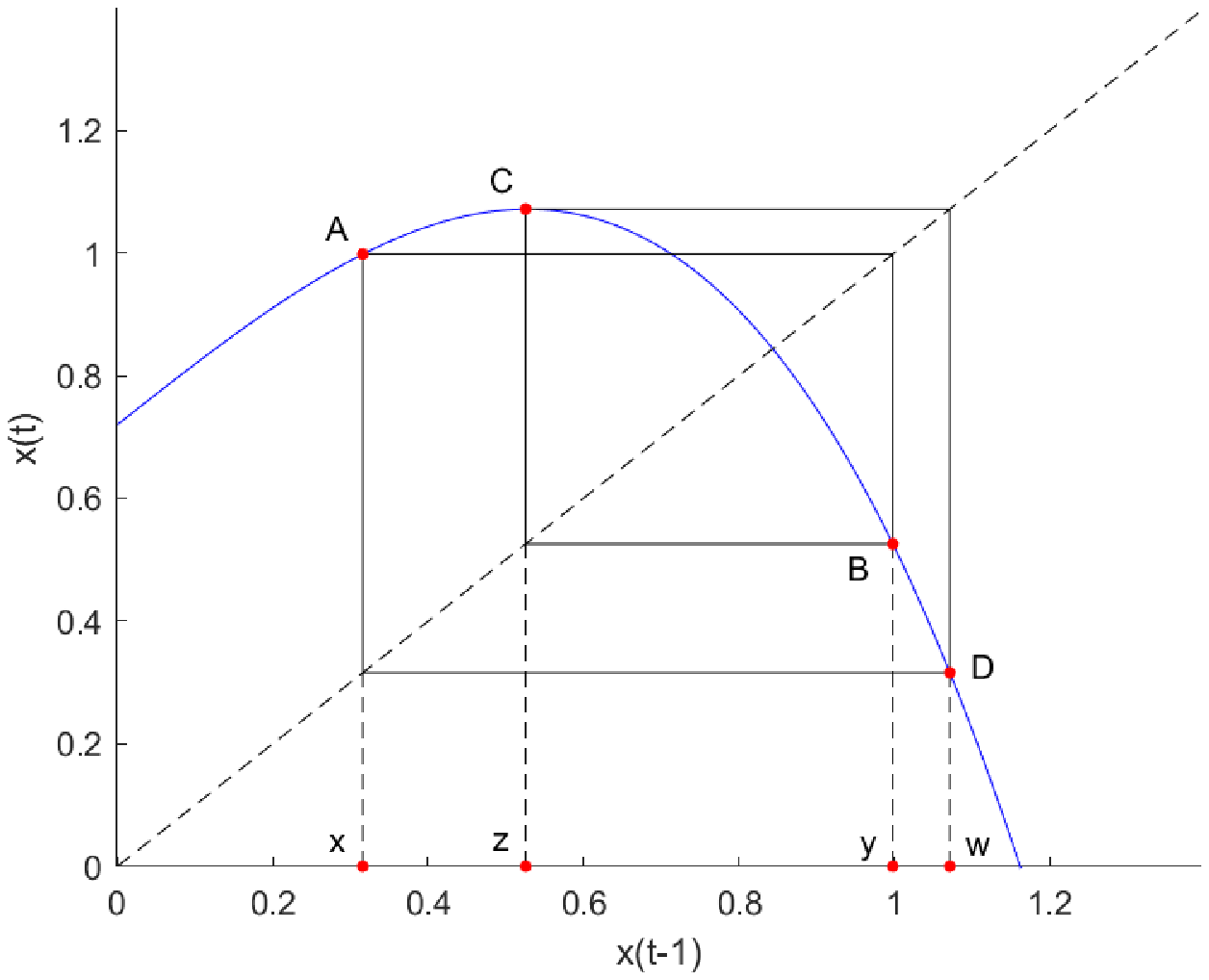}} 
    \subfigure[simplified phase diagram.]{\includegraphics[width=0.49\textwidth]{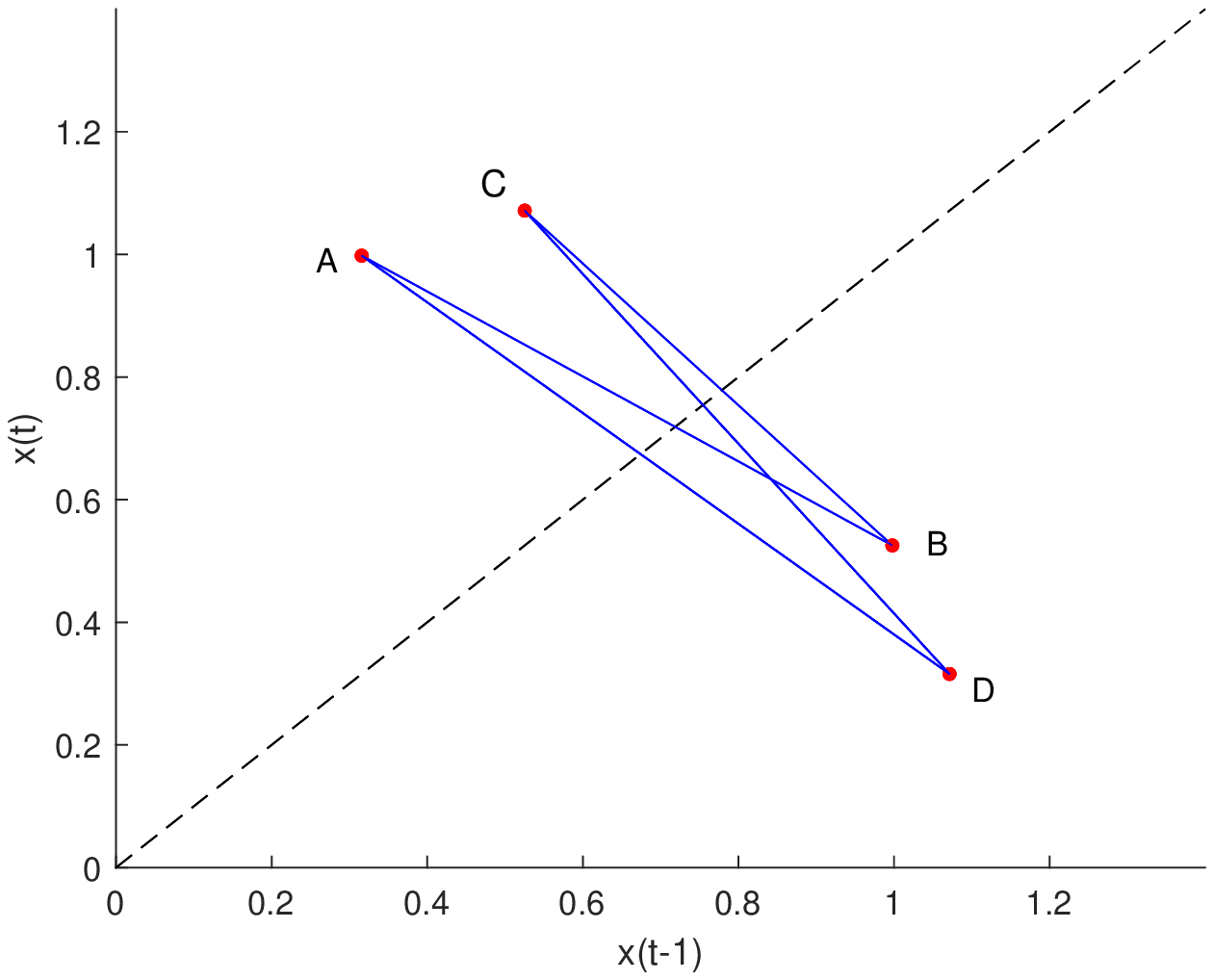}} \\
    
  \caption{The unique stable 4-cycle in Model 1 with $e=0.6$ and $f=1.2$.}
    \label{fig:4-cycle-model1}
\end{figure}

Figure \ref{fig:4-cycle-model1} (a) depicts the phase diagram of the unique 4-cycle in Model 1 with $e=0.6$ and $f=1.2$. Since $e^2f^3=0.62208\in (0.5794754859, 0.6673871142)$, this unique 4-cycle in Model 1 is asymptotically stable according to Theorem \ref{thm:mod1-4cycle}. Actually, the horizontal coordinates of $A,B,C,D$, i.e., $x,y,z,w$, are the four points in the 4-cycle orbit. For the sake of simplicity, we connect $A,B,C,D$ with lines and use the simplified phase diagram as Figure \ref{fig:4-cycle-model1} (b) to demonstrate periodic solutions in the rest of this paper.

Furthermore, by using the same approach as we computed the magnitude of the 2-cycle orbit, we conclude that if a 4-cycle $x\mapsto y\mapsto z\mapsto w\mapsto x$ exists in Model 1, its magnitude measure equals to
$$d=\frac{4}{f}\left(\frac{3\sqrt[3]{8-4e^2f^3+4(e^4f^6-4e^2f^3)^{1/2}}}{2}+\frac{6}{\sqrt[3]{8-4e^2f^3+4(e^4f^6-4e^2f^3)^{1/2}}}+2\right),$$
which is twice as large as that of the 2-cycle orbit.

\begin{figure}[htbp]
    \centering
    \includegraphics[width=8cm]{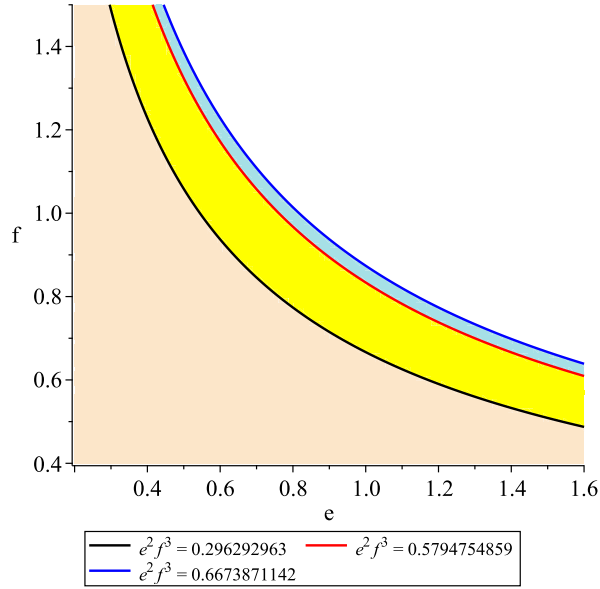}
    \caption{The parameter plane of Model 1. The light blue, yellow, and light orange regions are the parameter regions for the stability of the 4-cycle orbit, the 2-cycle orbit, and the equilibrium, respectively.}
    \label{fig:par-space-24cycle-model1}
\end{figure}

The parameter plane of Model 1 is shown in Figure \ref{fig:par-space-24cycle-model1}. One can see that the parameter region for the stability of the unique equilibrium (2-cycle or 4-cycle orbit) constitutes a connected set. Moreover, the three regions for the stability of the equilibrium, 2-cycle, and 4-cycle adjoin without any gap. In the next subsection, one will find that the topological structure of the parameter space of Model 2 is much more complex than that of Model 1.

\begin{figure}[htbp]
    \centering
    \includegraphics[width=16cm]{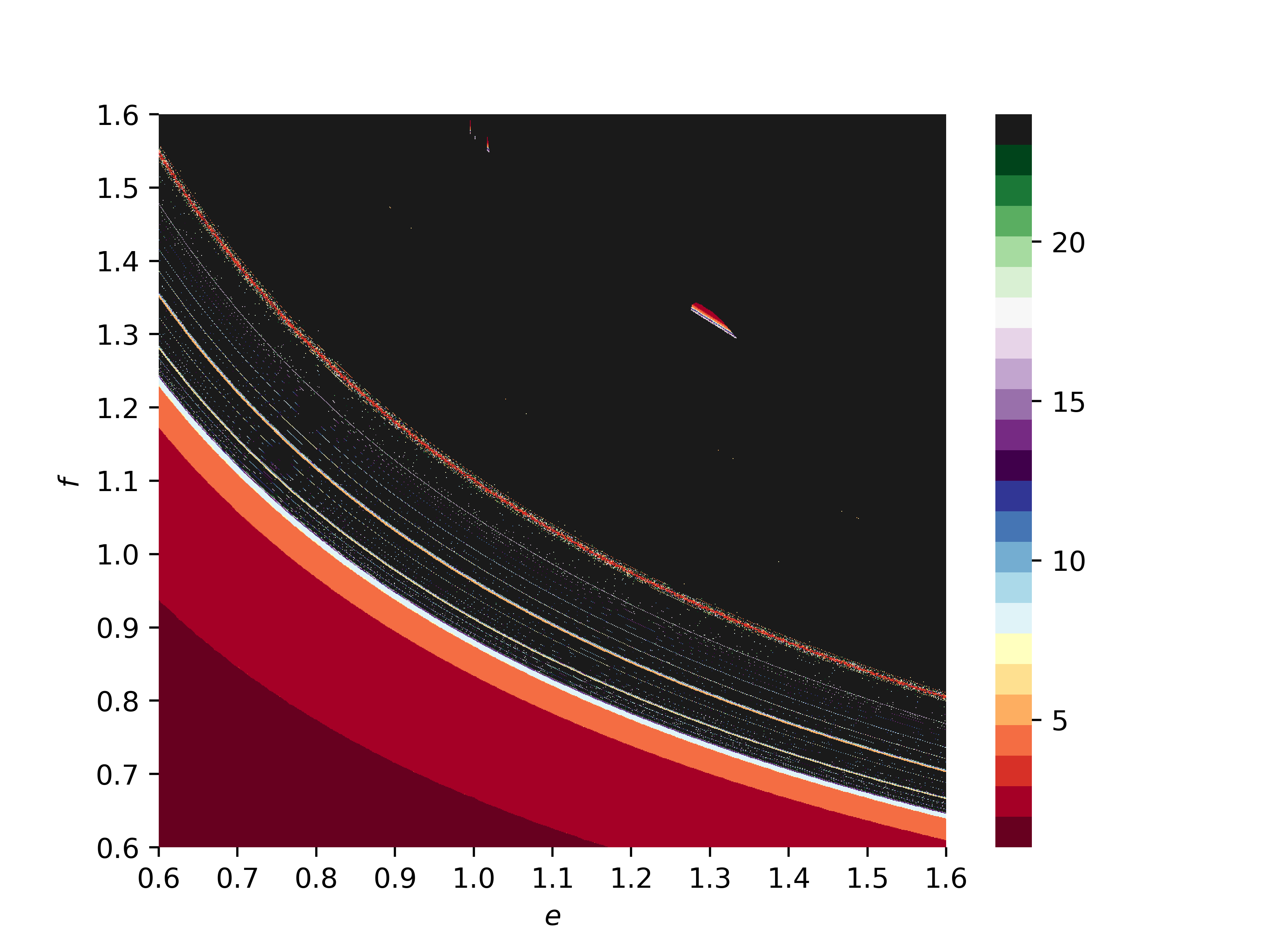}
    \caption{The two-dimensional bifurcation diagram of Model 1 for $(e,f)\in[0.6,1.6]\times[0.6,1.6]$. We choose $x(0)=1.0$ to be the initial state of the iterations.}
    \label{fig:bif-2d-mod1}
\end{figure}

Figure \ref{fig:bif-2d-mod1} depicts the two-dimensional bifurcation diagram of Model 1 for $(e,f)\in[0.6,1.6]\times[0.6,1.6]$. For additional information regarding two-dimensional bifurcation diagrams, readers can refer to \cite{Li2022C}. In the numerical simulations of Figure \ref{fig:bif-2d-mod1}, we set the initial state to be $x(0)=1.0$. Parameter points corresponding to periodic orbits with different orders are marked in different colors. For example, parameter points are colored in dark red if the order is just one (equilibria) and are marked in black if the order is greater than or equal to $24$ (complex trajectories). In the case that the order is greater than $24$, the black points may be viewed as the parameter values where complex dynamics such as chaos take place. Moreover, we also use black to mark those parameter points where the trajectories diverge to $\infty$. One can see that Figure \ref{fig:bif-2d-mod1} confirms the theoretical results presented in Figure \ref{fig:par-space-24cycle-model1}. 

\begin{figure}[htbp]
    \centering
    \includegraphics[width=14cm]{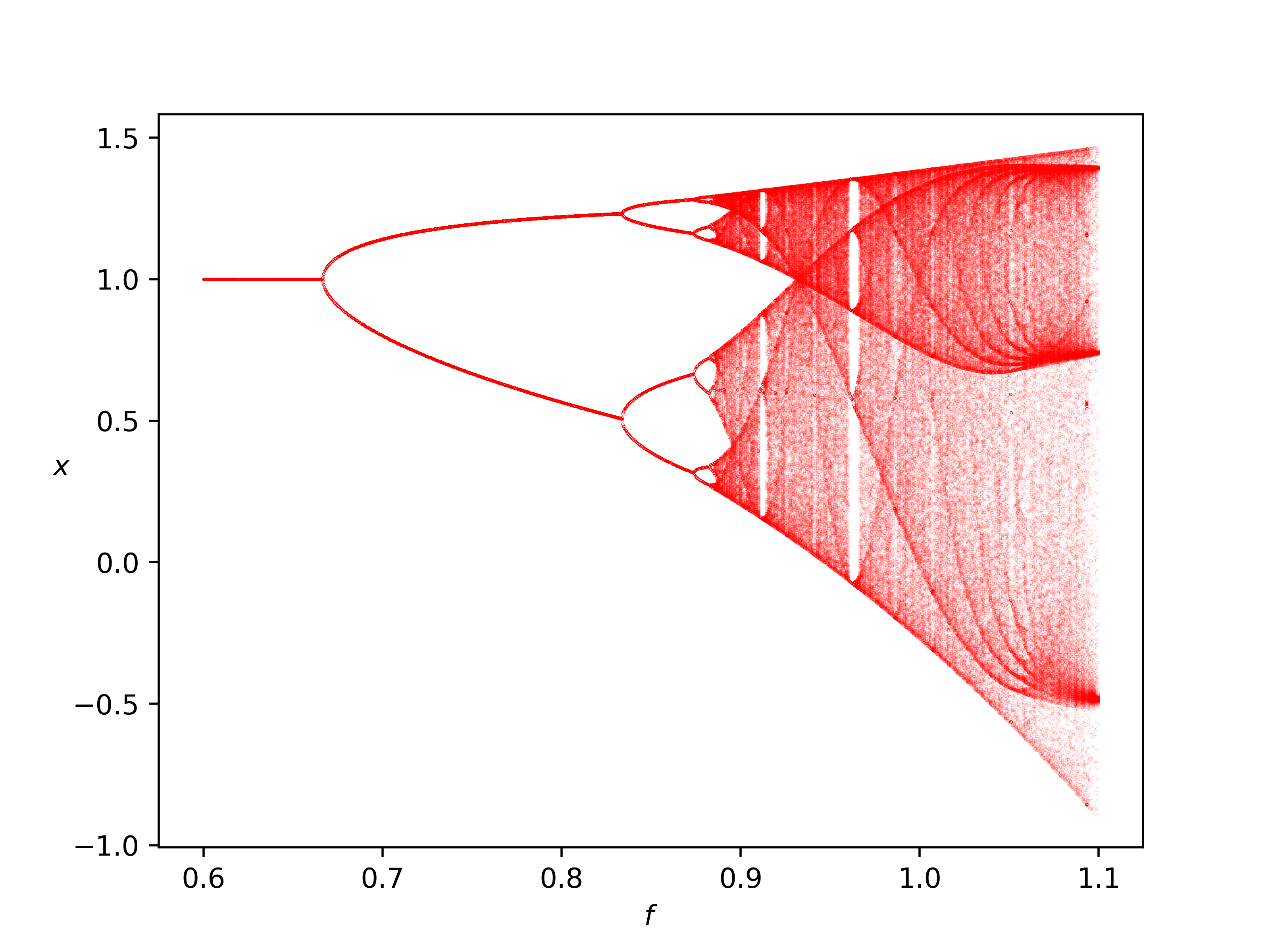}
    \caption{The one-dimensional bifurcation diagram of Model 1 with respect to $f$ by fixing $e=1.0$. We choose $x(0)=1.1$ to be the initial state of the iterations.}
    \label{fig:bifur_mod1}
\end{figure}

In Figure \ref{fig:bif-2d-mod1}, the transitions between different types of periodic orbits can also be observed. One can see that the equilibrium loses its stability through a series of period-doubling bifurcations as the value of $e$ or $f$ increases. For example, along the line of $e=1.0$, the unique stable equilibrium bifurcates into a stable 2-cycle orbit at $f =0.6665$, which further bifurcates into a 4-cycle orbit at $f =0.8339$. There is a stable 8-cycle orbit when $f \in (0.8744, 0.8826)$. Finally, chaotic dynamics take place if the value of $f$ is large enough. Additional details can be found in the one-dimensional bifurcation diagram presented in Figure \ref{fig:bifur_mod1}, where we fix $e=1.0$ and choose $x(0)=1.1$ to be the initial state of iterations.

\subsection{Model 2}
The formulation \eqref{eq:m2-iter} of Model 2 involves five parameters, which might be particularly complex for symbolic computations of searching periodic solutions. In what follows, we keep $K$ as the only parameter and assume that $a = 3.6$, $b = 2.4$, $c = 0.6$, and $d = 0.05$. This setting is meaningful and has been discussed by several economists, e.g., Puu \cite{Puu1995T},  Al-Hdaibat and others \cite{AlHdaibat2015O},  Matsumoto and Szidarovszky \cite{Matsumoto2022N}.

Let $x\mapsto y\mapsto x$ be a 2-cycle orbit. Hence, we have
\begin{equation}\label{eq:2-cycle-mod2}
	\left\{\begin{split}
		& y= x + K(3.6 - 4.8x + 1.8x^2 - 0.2x^3),\\
		& x= y + K(3.6 - 4.8y + 1.8y^2 - 0.2y^3),\\
		& x\neq y,\\
		& x>0, ~y>0, ~K>0.
	\end{split}\right.
\end{equation}
Furthermore, the following condition is required if the stability of the 2-cycle is considered.
$$| S(x)\cdot S(y)| <1,$$
where
\begin{equation}\label{eq:Sx}
S(x)=\frac{{\rm d}(x + K(3.6 - 4.8x + 1.8x^2 - 0.2x^3)}{{\rm d}x} =1-K(4.8-3.6x+0.6x^2).
\end{equation}

According to our computations, the following theorem is obtained.

\begin{theorem}\label{thm:mod2-2cycle}
In Model 2, the possible number of 2-cycle orbits is zero (no real solutions in system \eqref{eq:2-cycle-mod2}) or three (six real solutions in system \eqref{eq:2-cycle-mod2}). There exist three 2-cycle orbits if $K>5/3$. Moreover, two of them are stable if 
$$5/3<K<(5\sqrt{5}-5)/{3},$$ 
or approximately 
$$1.666666667<K<2.060113296.$$ 
\end{theorem}

To measure the magnitude of a 2-cycle orbit $x\mapsto y\mapsto x$, we also use $d=(x-y)^2+(y-x)^2$. The method of triangular decomposition permits us to decompose the solutions of 
\begin{equation*}
    \left\{\begin{split}
        &d=(x-y)^2+(y-x)^2,\\
        &y = x + K(3.6 - 4.8x + 1.8x^2 - 0.2x^3),\\
        &x = y + K(3.6 - 4.8y + 1.8y^2 - 0.2y^3)
    \end{split}\right.
\end{equation*}
into zeros of the following triangular systems.
\begin{equation*}
    \begin{split}
    \pset{T}_{31}=[\,&y-3, x-3, d\,],\\
    \pset{T}_{32}=[\,&y-x, x^2-6x+6, d\,],\\
    \pset{T}_{33}=[\,&y+x-6, Kx^2-6Kx+6K-10, Kd-24K-80\,],\\
    \pset{T}_{34}=[\,&5y+x^3K-9Kx^2+(24K-5)x-18K,\\ &K^2x^4-12K^2x^3+(51K^2-5K)x^2+(-90K^2+30K)x+54K^2-45K+25, \\
    &Kd-6K+10\,],\\
    \end{split}
\end{equation*}
where the last two polynomials $Kd-24K-80$ and $Kd-6K+10$ in {$\pset{T}_{33}$ and $\pset{T}_{34}$} are of our concern. We conclude that $d={(24K+80)}/{K}$ or $d={(6K-10)}/{K}$. One can see that two of the three 2-cycle orbits possess the same magnitude.

For a 3-cycle orbit $x\mapsto y\mapsto z \mapsto x$, we consider the system
\begin{equation}\label{eq:sys-3-cycle}
	\left\{\begin{split}
		& y= x + K(3.6 - 4.8x + 1.8x^2 - 0.2x^3),\\
		& z= y + K(3.6 - 4.8y + 1.8y^2 - 0.2y^3),\\
		& x= z + K(3.6 - 4.8z + 1.8z^2 - 0.2z^3),\\
		& {x\neq y,~x\neq z,}\\
		& x>0, ~y>0, ~z>0, ~K>0,
	\end{split}\right.
\end{equation}
as well as the stability condition 
\begin{equation}\label{eq:stab-cond3}
| S(x)\cdot S(y) \cdot S(z)| <1,	
\end{equation}
where 
$S(x)$ is given in \eqref{eq:Sx}. Based on a series of calculations, we have the following theorem. 


\begin{theorem}\label{thm:mod2-3cycle}

In Model 2, all possible cases for the number of (stable) 3-cycle orbits are listed in Table \ref{tab:3-cycles-model2}, where
\[m_1\approx 2.417401607,~ m_2\approx 2.434714456,~  m_3\approx 3.302953127,~m_4\approx 3.303122765.\]
Readers can refer to Remark \ref{rm:sp} to understand how these $m_i$ are obtained.

\begin{table}[H]
\caption{Numbers of (stable) 3-cycle orbits in Model 2}\label{tab:3-cycles-model2}
\centering
\begin{tabular}{|c|c|c|c|c|c|}
\hline $K\in$ & $(0,m_1)$ & $(m_1,m_2)$& $(m_2,m_3)$ & $(m_3,m_4)$& $(m_4,+\infty)$\\
\hline 3-cycles & 0 & 4 & 4 & 8 & 8\\
\hline Stable 3-cycles & 0 & 2 & 0 & 2 & 0\\
 \hline
\end{tabular}
\end{table}
\end{theorem}

\begin{remark}\label{rm:sp}
As aforementioned, the border polynomial plays an important role. However, one can derive that the properties of the border polynomial reported in Lemma \ref{lem:main} will retain if we use the squarefree part of the border polynomial. The squarefree part $SP$ of the border polynomial of \eqref{eq:sys-3-cycle}+\eqref{eq:stab-cond3} is simpler, which is given in Appendix. In Theorem \ref{thm:mod2-3cycle}, $m_1,\ldots,m_4$ are the real roots of $SP$. A rigorous style of writing Theorem \ref{thm:mod2-3cycle} is to express the conditions using factors in $SP$. However, this would be quite tedious. Since only one parameter, i.e., $K$, is involved in $SP$, the regions divided by zeros of $SP$ are indeed intervals and can be approximately described as in Theorem \ref{thm:mod2-3cycle}. However, it should be noticed that the values of $m_1,\ldots,m_4$ can be made arbitrarily accurate if we want because the exact expression of $SP$ has already been obtained.
\end{remark}

%

\begin{figure}[htbp]
    \centering
    \includegraphics[width=10cm]{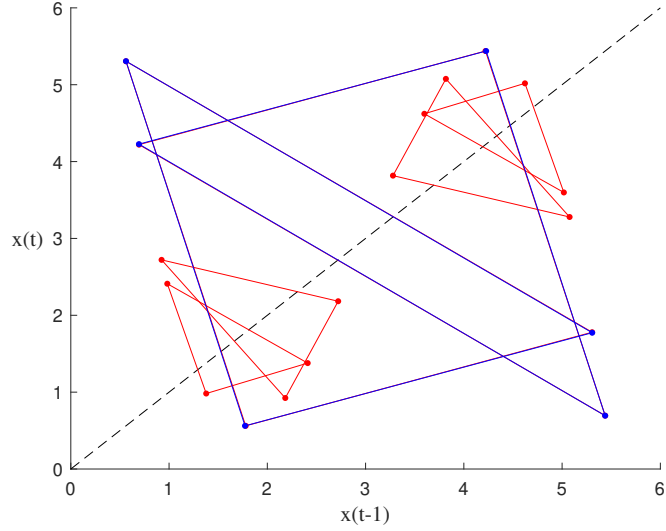}
    \caption{The 3-cycle orbits in Model 2 with $K=3.303$. The 6 unstable cycles are marked in red, while the 2 stable ones are marked in blue.}
    \label{fig:3-cycle-mod2}
\end{figure}

Figure \ref{fig:3-cycle-mod2} depicts all the 3-cycle orbits in Model 2 with $K=3.303\in (m_3,m_4)$, where the $6$ unstable cycles are marked in red and the $2$ stable cycles are marked in blue. It is worth noting that two of the unstable 3-cycle orbits in red almost coincide with the stable ones in blue, but they are different. We should underline that this dynamic phenomenon, derived by symbolic computations, may be too subtle to observe through numerical simulations.

\begin{figure}[htbp]
    \centering
    \includegraphics[width=7cm]{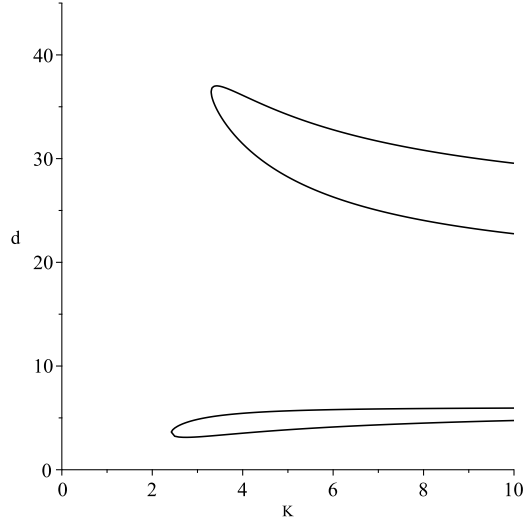}
    \caption{The magnitude $d$ of the possible 3-cycle orbits in Model 2 as the variation of $K$.}
    \label{fig:Kd-3-cycle}
\end{figure}

Moreover, if measuring the magnitude of the 3-cycle orbit $x\mapsto y\mapsto z \mapsto x$ with $d=(x-y)^2+(y-z)^2+(z-x)^2$, then we have
\begin{align*}
&K^{4} d^{4}
+(
-54K^{4}
-90K^{3}) d^{3}
+(972K^{4}
+2700K^{3}
+1800K^{2}) d^{2}\\
&~~+(
-6696K^{4}
-19440K^{3}
-5400K^{2}
+27000K) d\\
&~~+15552K^{4}
+38880K^{3}
-32400K^{2}
-162000K
+270000=0.
\end{align*}
The above condition on $K$ and $d$ is plotted in Figure \ref{fig:Kd-3-cycle}.

Similarly, we analyze the 4-cycle and 5-cycle orbits in Model 2, and report the obtained results in the sequel. 
\begin{theorem}\label{thm:mod2-4cycle}
In Model 2, all possible cases for the number of (stable) 4-cycle orbits are given in Table \ref{tab:4-cycles-mod2}, where
\begin{align*}
&m_1\approx 2.060113296,~m_2\approx 2.146719591,~ m_3\approx 2.579725065,~ m_4\approx 2.581385365,~ m_5\approx 3.062775154,\\
&m_6\approx 3.070194019,~m_7\approx 3.279225134,~ m_8\approx 3.279260335,~ m_9\approx 3.319881360,~ m_{10}\approx 3.319889702.   
\end{align*}
Readers can refer to Remark \ref{rm:sp} to understand how these $m_i$ are obtained.

\begin{table}[H]
\caption{Numbers of (stable) 4-cycle orbits in Model 2}\label{tab:4-cycles-mod2}
\centering
\begin{tabular}{|c|c|c|c|c|c|c|}
\hline $K\in$ & $(0,m_1)$ & $(m_1,m_2)$& $(m_2,m_3)$ & $(m_3,m_4)$& $(m_4,m_5)$ & $(m_5,m_6)$ \\
\hline 4-cycles & 0 & 2 & 2 & 6 & 6&10\\
\hline Stable 4-cycles & 0 & 2 & 0 & 2 & 0&2\\
 \hline
 \hline $K\in$& $(m_6,m_7)$  & $(m_7,m_8)$&$(m_8,m_9)$ & $(m_9,m_{10})$& $(m_{10},+\infty)$ &\\
\hline 4-cycles & 10 & 14 & 14 & 18 & 18&\\
\hline Stable 4-cycles & 0 & 2 & 0 & 2 & 0&\\
\hline
\end{tabular}
\end{table}
\end{theorem}

\begin{figure}[htbp]
    \centering
    \includegraphics[width=10cm]{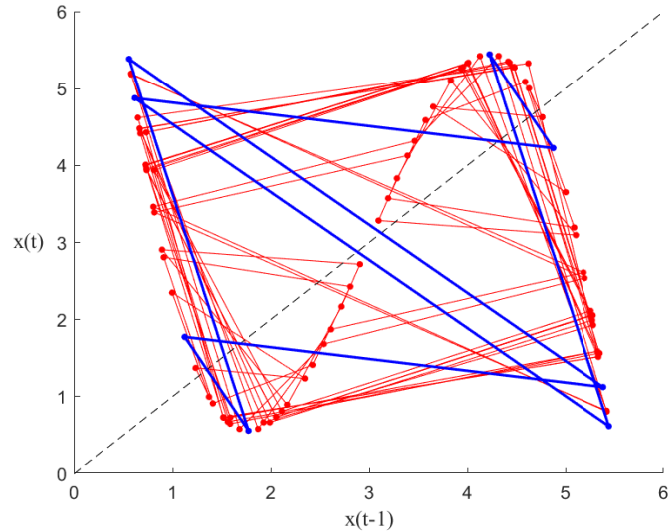}
    \caption{The 4-cycle orbits in Model 2 with $K=3.319885$. The 16 unstable cycles are marked in red, while the 2 stable ones are marked in blue.}
    \label{fig:4-cycle-mod2}
\end{figure}

In Figure \ref{fig:4-cycle-mod2}, we show all the 4-cycle orbits in Model 2 with $K=3.319885\in(m_9,m_{10})$, where the $16$ unstable cycles are marked in red and the $2$ stable ones are marked in blue. If we measure the magnitude of the 4-cycle orbit $x\mapsto y\mapsto z \mapsto w \mapsto x$ with $d=(x-y)^2+(y-z)^2+(z-w)^2+(w-x)^2$, then $d$ must satisfy one of the following equations.
\begin{align*}
& Kd-12K+20=0,\\
& Kd-48K-160=0,\\
& K^2d^2+(-36K^2-60K)d+288K^2+960K+1600=0,\\
& C_4(K,d)=0,
\end{align*}
where $C_4(K,d)$ is a complex polynomial given in Appendix.

%
%

\begin{theorem}\label{thm:mod2-5cycle}
In Model 2, all possible cases for the number of (stable) 5-cycle orbits are listed in Table \ref{tab:5-cycles-mod2}, where
\begin{align*}
&m_1\approx 2.323208379,~ m_2\approx  2.326320457,~  m_3\approx 2.509741151,~ m_4\approx 2.510528490,~ \\
&m_5\approx 2.632885028,~m_6\approx 2.633089005,~m_7\approx 2.997641294,~ m_8\approx 2.997736262,~ \\
&m_9\approx 3.113029799,~ m_{10}\approx 3.113069634,~m_{11}\approx 3.197332995,~m_{12}\approx 3.197354147,~\\
&m_{13}\approx 3.219425160,~ m_{14}\approx 3.219440784,~  m_{15}\approx 3.269613400,~m_{16}\approx 3.269618202,~\\
&m_{17}\approx 3.288059620,~ m_{18}\approx 3.288062995,~m_{19}\approx 3.314977518,~ m_{20}\approx 3.314978815,~\\
&m_{21}\approx 3.324008184,~m_{22}\approx 3.324008826,~  m_{23}\approx 3.332961824,~ m_{24}\approx 3.332961850.
\end{align*}
Readers can refer to Remark \ref{rm:sp} to understand how these $m_i$ are obtained.

\begin{table}[H]
\caption{Numbers of (stable) 5-cycle orbits in Model 2}\label{tab:5-cycles-mod2}
\centering
\begin{tabular}{|c|c|c|c| c| c|}
\hline $K\in$ & $(0,m_1)$ & $(m_1,m_2)$& $(m_2,m_3)$ & $(m_3,m_4)$& $(m_4,m_5)$ \\
\hline 5-cycles & 0 & 4 & 4 & 8 & 8 \\
\hline Stable 5-cycles & 0 & 2 & 0 & 2 & 0\\
 \hline
 \hline $K\in$ & $(m_5,m_6)$ & $(m_6,m_7)$  & $(m_7,m_8)$&$(m_8,m_9)$ & $(m_9,m_{10})$\\
\hline 5-cycles & 12 & 12 & 16 & 16 & 20\\
\hline Stable 5-cycles &2 & 0 & 2 & 0 & 2\\
\hline
 \hline $K\in$ & $(m_{10},m_{11})$ & $(m_{11},m_{12})$ & $(m_{12},m_{13})$ & $(m_{13},m_{14})$ & $(m_{14},m_{15})$\\
\hline 5-cycles &20 &24 & 24 & 28 & 28 \\
\hline Stable 5-cycles &0 &2 & 0 & 2 & 0 \\
\hline
 \hline $K\in$ &$(m_{15},m_{16})$ &$(m_{16},m_{17})$ & $(m_{17},m_{18})$ & $(m_{18},m_{19})$ & $(m_{19},m_{20})$\\
 \hline 5-cycles &32 &32 &36 & 36 & 40 \\
\hline Stable 5-cycles &2 &0 &2 & 0 & 2\\
\hline
\hline $K\in$ & $(m_{20},m_{21})$ & $(m_{21},m_{22})$ & $(m_{22},m_{23})$ & $(m_{23},m_{24})$& $(m_{24},+\infty)$  \\
 \hline 5-cycles & 40 & 44 & 44&48 & 48 \\
\hline Stable 5-cycles  & 0 & 2 & 0&2 & 0 \\
\hline
\end{tabular}
\end{table}

\end{theorem}

\begin{figure}[htbp]
    \centering
    \includegraphics[width=10cm]{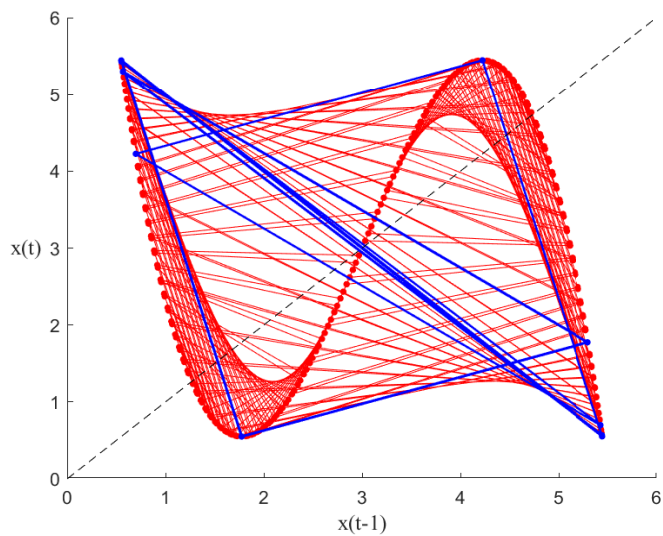}
    \caption{The 5-cycle orbits in Model 2 with $K=3.33296183$. The 46 unstable cycles are marked in red, while the 2 stable ones are marked in blue.}
    \label{fig:5-cycle-mod2}
\end{figure}

In Figure \ref{fig:5-cycle-mod2}, we plot all possible 5-cycle orbits in Model 2 with $K=3.33296183\in (m_{23},m_{24})$, where the $46$ unstable cycles are marked in red and the $2$ stable ones are marked in blue. 

By Theorems \ref{thm:mod2-3cycle}, \ref{thm:mod2-4cycle} and \ref{thm:mod2-5cycle}, one can see the parameter space of Model 2 is quite different from that of Model 1 in the sense that the stability regions for the 3-cycle, 4-cycle and 5-cycle orbits are disconnected sets formed by many disjoint portions. Therefore, the topological structures of the regions for stable periodic orbits in Model 2 are much more complex than those in Model 1. This may be because the inverse demand function of Model 2 has an inflection point. However, the following observations of Model 2 are similar to Model 1. Theorem \ref{thm:mod2-2cycle} shows that the stability region for the 2-cycles is a connected interval. In Model 2, the right boundary of the stability region for the 2-cycles is the same as the left boundary of the stability region for the 4-cycles. When $a = 3.6$, $b = 2.4$, $c = 0.6$, and $d = 0.05$, by Theorem \ref{thm:mod2-stab} we know that Model 2 has stable equilibria if $K\in (0,5/3)$, which adjoins the stability region for the 2-cycles. Moreover, in Model 2, the stability regions for cycles with distinct periods may not intersect with each other, which means that multistability might only arise among periodic orbits with the same period.

\begin{figure}[htbp]
    \centering
    \includegraphics[width=16cm]{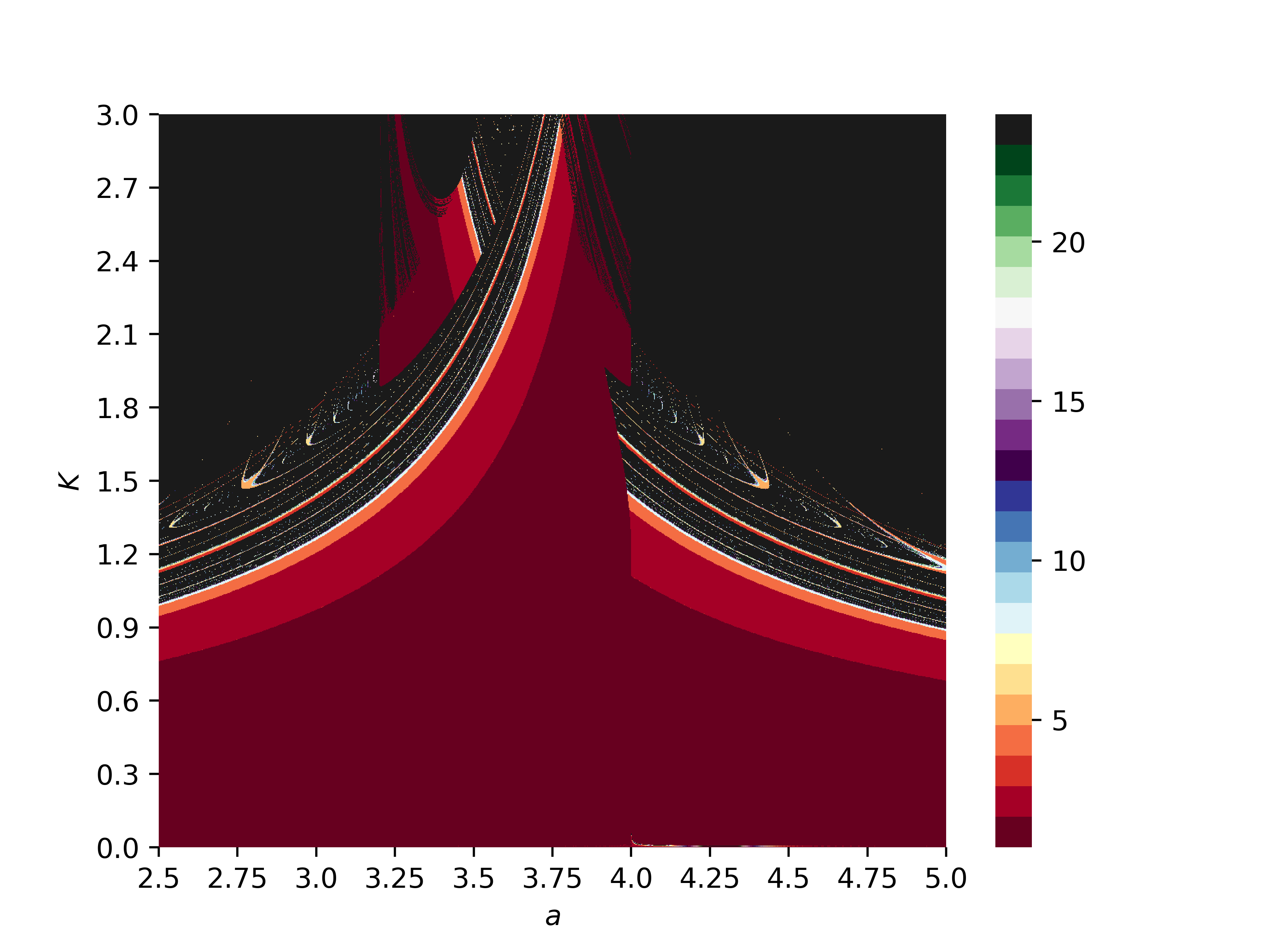}
    \caption{The two-dimensional bifurcation diagram of Model 2 for $(a,K)\in[2.5,5.0]\times[0.0,3.0]$. We fix the parameters  $b=2.4$, $c=0.6$, $d=0.05$, and choose $x(0)=1.0$ to be the initial state of the iterations.}
    \label{fig:bif-2d-mod2}
\end{figure}

Figure \ref{fig:bif-2d-mod2} depicts the two-dimensional bifurcation diagram of Model 2 for $(a,K)\in[2.5,5.0]\times [0.0,3.0]$. We fix the parameters $b=2.4$, $c=0.6$, $d=0.05$, and set the initial state to be $x(0)=1.0$. Similarly, we use different colors to mark parameter points corresponding to trajectories with different periods. Parameter points are marked in black if the corresponding orbits have orders greater than 24. Furthermore, we also use black to mark the parameter points where the trajectories diverge. One can see that Figure \ref{fig:bif-2d-mod2} confirms the theoretical results reported in Theorem \ref{thm:mod2-stab}. However, Figure \ref{fig:bif-2d-mod2} generated by numerical simulations is not accurate compared to Figure \ref{fig:par-plane-mod2} based on symbolic computations.

\begin{figure}[htbp]
  \centering
    \subfigure[$x(0)=1.0$.]{\includegraphics[width=0.49\textwidth]{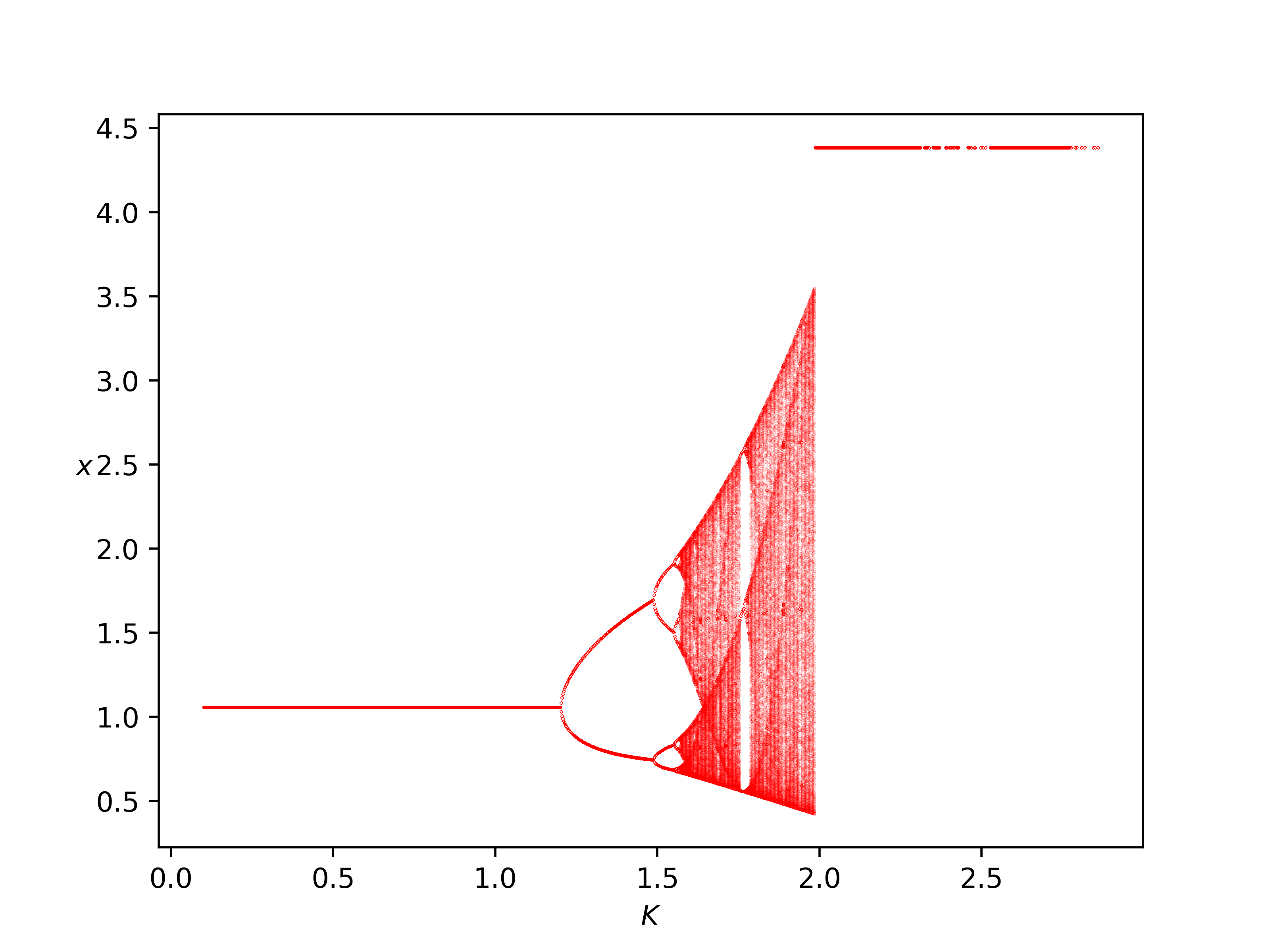}} 
    \subfigure[$x(0)=4.0$.]{\includegraphics[width=0.49\textwidth]{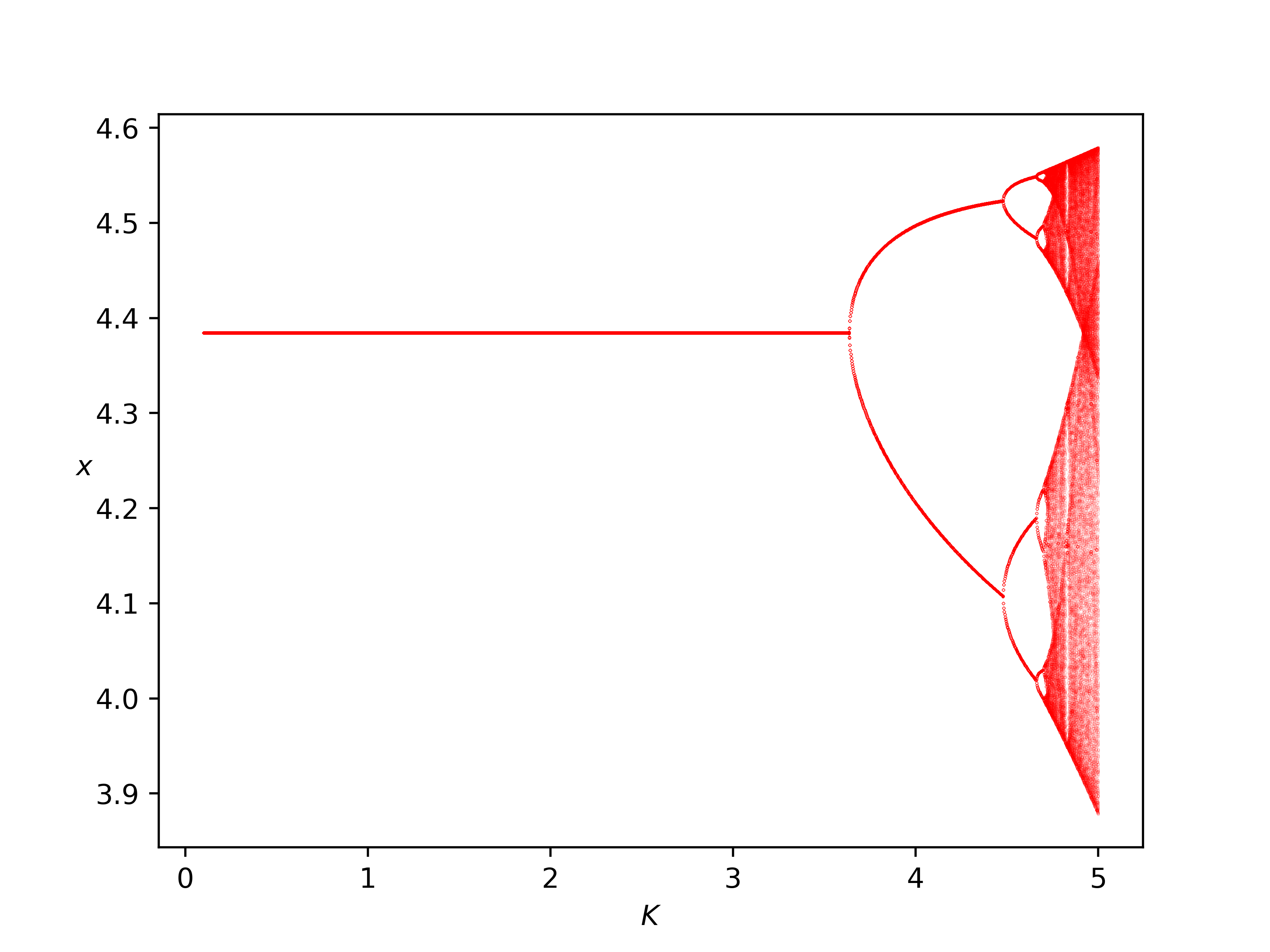}} \\
  \caption{The one-dimensional bifurcation diagrams of Model 2 with respect to $K$ by fixing $a=3.3$, $b=2.4$, $c=0.6$, and $d=0.05$.}
    \label{fig:1d-bif-K}
\end{figure}

Figure \ref{fig:1d-bif-K} depicts the one-dimensional bifurcation diagrams of Model 2 with respect to $K$ by fixing $a=3.3$, $b=2.4$, $c=0.6$, and $d=0.05$. The bifurcation diagrams are different if the selected initial states of the iterations are distinct. For example, in Figure \ref{fig:1d-bif-K} (a) and (b), the initial states are selected to be $x(0)=1.0$ and $x(0)=4.0$, respectively. The difference may be because two stable equilibria exist when $K$ is relatively small and distinct initial states approach distinct equilibria. As shown by Figure \ref{fig:1d-bif-K} (a), the trajectory converges to $1.058$ when $K<1.1996$ and converges to $4.384$ when $K>1.9874$. In Figure \ref{fig:1d-bif-K}, the occurrence of period-doubling bifurcations can also be observed.

\begin{figure}[htbp]
  \centering
    \subfigure[$x(0)=1.0$.]{\includegraphics[width=0.49\textwidth]{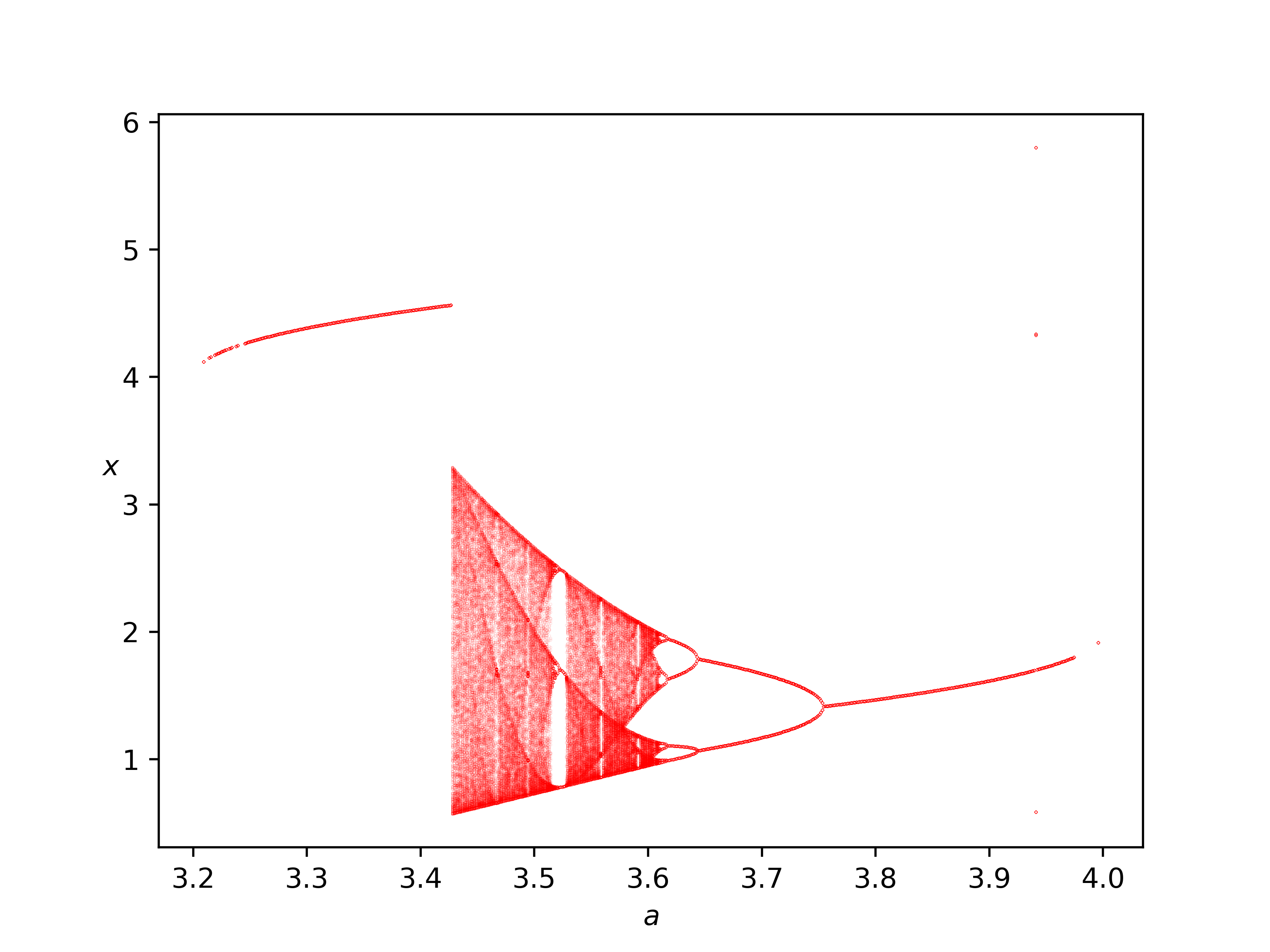}} 
    \subfigure[$x(0)=4.0$.]{\includegraphics[width=0.49\textwidth]{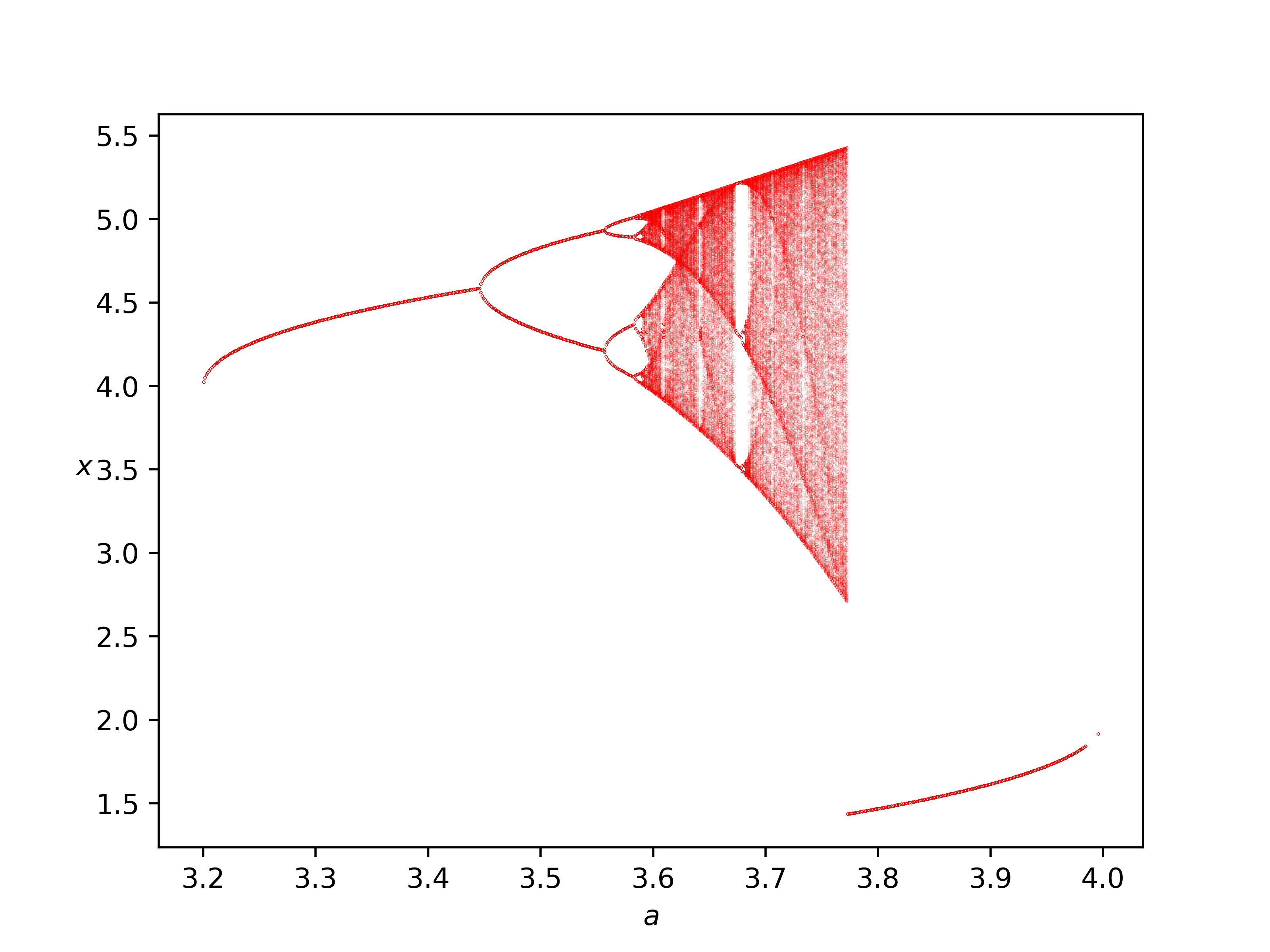}} \\
  \caption{The one-dimensional bifurcation diagrams of Model 2 with respect to $a$ by fixing $K=2.2$, $b=2.4$, $c=0.6$, and $d=0.05$.}
    \label{fig:1d-bif-a}
\end{figure}

Figure \ref{fig:1d-bif-a} depicts the one-dimensional bifurcation diagrams of Model 2 with respect to $a$ by fixing $K=2.2$, $b=2.4$, $c=0.6$, and $d=0.05$. In Figure \ref{fig:1d-bif-a} (a) and (b), the initial states of the iterations are selected to be $x(0)=1.0$ and $x(0)=4.0$, respectively. Similarly, the two bifurcation diagrams are different because of the selection of distinct initial states. Furthermore, pitchfork bifurcations can be observed in Figure \ref{fig:1d-bif-a}, where the number of stable equilibria changes from one to zero.

In Model 2, two stable equilibria may coexist (see the blue-gray region in Figure \ref{fig:par-plane-mod2}). The equilibrium selection problem is interesting. The final outcome of the iterations depends not only on the values of the parameters but also on the starting conditions of the game. According to our numerical simulations of Model 2, the basins of attraction of coexisting equilibria have complicated structures. For example, by fixing $K=0.5$, $a=3.5$, $b=2.4$, $c=0.6$, $d=0.05$, we have two stable equilibria $E_1=1.19$ and $E_2=4.64$. The basin of $E_1$ is 
$$\pset{B}(E_1)=(0,3.168)\cup (6.518,7.577)\cup (7.745,7.781)\cup (7.786,7.789),$$
while that of $E_2$ is 
$$\pset{B}(E_2)=(3.168,6.518)\cup (7.577,7.745)\cup (7.781,7.786).$$
Furthermore, when the initial state $x(0)>7.786$, the trajectory will not converge to any of the two stable equilibria but diverge to $+\infty$. Take $K=1$ and $a=4$ as the other example. If the other parameters keep unchanged, i.e., $b=2.4$, $c=0.6$, and $d=0.05$, there are two stable equilibria $E_1=4.99$ and $E_2=1.99$. Our simulations show that the basins of these two equilibria are
$$\pset{B}(E_1)=(0,0.807)\cup (2.0,6.192)\cup (6.431,6.647)\cup (6.653,6.659),$$ 
and
$$\pset{B}(E_2)=(0.807,2.0)\cup (6.192,6.431)\cup (6.647,6.653),$$
respectively. The escape set is $(6.659,+\infty)$, where the trajectory diverges.
In short, in Model 2, the basins of the two stable equilibria are disconnected sets and have complex topological structures.

%

\section{Chaotic Dynamics}

In the bifurcation diagrams (Figures \ref{fig:bifur_mod1} and \ref{fig:1d-bif-K}), one can observe that the dynamics of the two considered models transition to chaos through period-doubling bifurcations as the adjustment speed increases. From an economic point of view, if chaos appears, the pattern behind output and profits is nearly impossible to learn even for completely rational players. Therefore, it is extremely hard for a firm to handle a chaotic economy, where no market rules could be discovered and followed.

In this section, we rigorously prove the existence of chaos for the two models. The following famous lemma was first derived by Li and Yorke \cite{Li1975P}, which is mathematically deep and facilitates the exploration of complicated dynamics arising in one-dimensional discrete dynamical systems.

\begin{lemma}\label{lem:p3chaos}
Let $I$ be an interval of real numbers, and let $F: I\rightarrow \field{R}$ be a continuous function. Assume that there exists a point $x\in I$ such that 
\begin{equation}\label{eq:p3}
	F^3(x)\leq x <F(x)<F^2(x)~~~\text{or}~~~~F^3(x)\geq x >F(x)>F^2(x),
\end{equation}
then the following two statements are true.
\begin{enumerate}
	\item For each $k\in \{1,2,\ldots\}$, there is a point $p_k\in I$ with period $k$, i.e., $F^k(p_k)=p_k$, and $F^i(p_k)\neq p_k$ for $1\leq i <k$.
	
	\item There is an uncountable set $S\subset I$ (containing no periodic points), which satisfies the following conditions:
	\begin{enumerate}
		\item for any $p,q\in S$ with $p\neq q$,
		\begin{equation}\label{eq:away2p}
			\limsup_{n\rightarrow\infty}|F^n(p)-F^n(q)|>0,
		\end{equation}
				and
		\begin{equation}\label{eq:near2p}
			\liminf_{n\rightarrow\infty}|F^n(p)-F^n(q)|=0;
		\end{equation}
		$$$$		
		\item for every point $p\in S$ and every periodic point $q\in I$,
		\begin{equation}\label{eq:awaycircle}
			\limsup_{n\rightarrow\infty}|F^n(p)-F^n(q)|>0.
		\end{equation}

	\end{enumerate}
\end{enumerate}
\end{lemma}

\begin{remark}
Eq.\ \eqref{eq:near2p} means that every trajectory in $S$ can wander arbitrarily close to every other. 
However, by \eqref{eq:away2p} we know that no matter how close two distinct trajectories in $S$ may come to each other, they must eventually wander away.
Furthermore, by \eqref{eq:awaycircle} it is clear that every trajectory in $S$ goes away from any periodic orbit in $I$. If the two statements in the above lemma are both satisfied, we say that there exist chaotic dynamics or chaos in the sense of Li-Yorke.
\end{remark}


Therefore, we can conclude that ``period three implies chaos'' for one-dimensional discrete dynamical systems. In Section 4, we have rigorously derived the existence of 3-cycle orbits in Model 2 if $K>2.417401607$, which proves that chaos would arise for an uncountable set of initial states in the sense of Li-Yorke. 

But in Model 1, we have proved that there are no solutions with period three. However, it can not be concluded that there exist no chaotic trajectories since the existence of period three is not a necessary but only a sufficient condition of chaos. In \cite{Marotto1978S}, Marotto indicated that the existence of snapback repellers also implies chaos for general $n$-dimensional systems. However, Li and Chen \cite{Li2003O} pointed out that Marotto's original definition of snapback repeller may result in an insufficiency, and proposed the Marotto-Li-Chen Theorem. Thus, we give the following lemma for one-dimensional systems by simplifying the Marotto-Li-Chen Theorem. Readers can refer to \cite{Huang2019A} for additional details.



\begin{lemma}\label{lem:marotto}
Let $I$ be an interval of real numbers, and let $F: I\rightarrow \field{R}$ be a differentiable function. Assume that
\begin{enumerate}
    \item $x\in I$ is an equilibrium, i.e., $F(x)=x$;
    \item there exists a close interval $S\subset I$ such that $x$ is an inner point of $S$, and the derivative of $F$ has the absolute value greater than $1$ at every point $p\in S$, i.e., $|F'(p)|>1$;
    \item for some integer $m>1$, there exists a point $y\in S$ such that $y\neq x$, $F^m(y)=x$, and $F'(F^k(y))\neq 0$ for all $1\leq k\leq m$.
\end{enumerate}
Then the system $x(t+1)=F(x(t))$ is chaotic in the sense of Li-Yorke.
\end{lemma}


For Model 1, we have $F(x)=x+f(e-x^3)$ and $F'(x)=1-3fx^2$. Then $|F'(x)|>1$ and $x>0$ imply that $x>\sqrt{\frac{2}{3f}}$. Thus, if we can find $x,y$ with $x\neq y$ such that both $|F'(x)|>1$ and $|F'(y)|>1$ are satisfied, then there must exist one closed interval $S$ containing $x,y$ as inner points. In such a case, it is obvious that $|F'(p)|>1$ for every point $p\in S$. Naturally, we start from $m=2$ to verify the conditions of Lemma \ref{lem:marotto} by counting real solutions of the following system.

\begin{equation*}
	\left\{\begin{split}
		& x=x + f(e-x^3),\\
		& x=F^2(y)=y+f(e-y^3)+f(e-(y+f(e-y^3))^3),\\
		& |1-3fx^2|>1,\\
		& |1-3fy^2|>1,\\
		& |1-3f(y+f(e-y^3))^2|\neq 0,\\
		& x\neq y,\\
		& x>0, ~y>0, ~e>0, ~f>0.
	\end{split}\right.
\end{equation*}
The technique introduced in Remark \ref{rm:linearization} should be conducted first to transform the above system into a univariate one. According to our calculations, the above system has at least one real solution if and only if $8/27<e^2f^3<64/27$. Therefore, we conclude that Model 1 is chaotic in the sense of Li-Yorke provided that $8/27<e^2f^3<64/27$.


\section{Concluding Remarks}

It is known that a monopoly may exhibit complex dynamics such as periodic orbits and chaos although it is the simplest oligopoly. In this study, we investigated two monopoly models with gradient mechanisms, where the monopolists are knowledgeable firms. The two models are distinct mainly in their inverse demand functions. Model 1 uses the inverse demand function of Naimzada and Ricchiuti \cite{Naimzada2008C}, while Model 2 employs that of Puu \cite{Puu1995T}. Different from widely applied numerical methods such as numerical simulations and bifurcation continuation approaches, symbolic methods were applied in this paper to analyze the local stability, periodic solutions, and even chaotic dynamics. Numerical methods have some shortcomings, e.g., the computations may encounter the problem of instability, which makes the results completely useless. In comparison, symbolic computations are exact, thus the obtained results can be used to rigorously prove economic theorems in some sense. 

By reproving the already-known results (Proposition \ref{prop:m1-stab}) of the local stability and bifurcations of Model 1, we explained in detail how our symbolic approach works. Afterward, the analysis of the stability and bifurcations of Model 2 was conducted based on this approach. We acquired the complete conditions of the local stability and bifurcations of Model 2 for the first time (see Theorem \ref{thm:mod2-stab}). In Figure \ref{fig:par-plane-mod2}, it was observed that Model 2 behaves quite differently from typical oligopoly models with gradient mechanisms. For example, even if the adjustment speed $K$ is quite large, there always exist some values of $a$ (the difference between the initial commodity price and the initial marginal cost) such that Model 2 has a stable equilibrium. Moreover, Model 2 may go from instability to stability and then back to instability twice as the value of $a$ increases.

From an economic point of view, the study of periodic solutions is of practical importance. Under the assumption of bounded rationality, firms can not learn the pattern behind output and profits if periodic dynamics take place. For the two models, we explored the periodic solutions with lower orders as well as their local stability. Differences between the two models were found, e.g., 3-cycle orbits exist in Model 2 but not in Model 1. In Model 1, the parameter region for the stability of the periodic solution with a fixed order constitutes a connected set. In Model 2, however, the stability regions for the 3-cycle, 4-cycle, and 5-cycle orbits are disconnected sets formed by many disjoint portions. In other words, the topological structures of the regions for stable periodic orbits in Model 2 are much more complex than those in Model 1. The above differences may be because the inverse demand function of Model 2 has an inflection point. According to the numerical simulations of Model 2, we found that the basins of the two stable equilibria are disconnected sets and also have complex topological structures. For a $n$-cycle orbit $p_1\mapsto p_2\mapsto \cdots p_n\mapsto p_1$, we defined the magnitude measure to be
$$d=(p_1-p_2)^2+(p_2-p_3)^2+\cdots+(p_{n-1}-p_n)^2+(p_n-p_1)^2.$$
For the two considered models, we analytically investigated the formulae for the magnitude of periodic orbits with lower orders.

Furthermore, it is extremely hard for a firm to handle an economy when chaos appears. In such a case, no market rules can be discovered and followed, and the pattern behind output and profits is nearly impossible to learn even for completely rational players. In the bifurcation diagrams of the two models, it seems that chaos occurs when the adjustment speed is large enough. We clarified this observation analytically. By virtue of the fact ``period three implies chaos'', we derived that Model 2 is chaotic in the sense of Li-Yorke by proving the existence of 3-cycle orbits. However, there are no 3-cycles in Model 1, but the Marotto-Li-Chen Theorem permitted us to prove the existence of chaos by finding snapback repellers. 

In this paper, we take the assumption of knowledgeable players, which means the enterprise has full information regarding the inverse demand function and can compute its marginal profit at any time. In the real world, however, it is more reasonable to assume players to be limited rather than knowledgeable. In this case, the enterprise does not know the form of the inverse demand function, but possesses the values of output and price only in the past periods and estimates its marginal profit with a simple difference formula. The investigation of the dynamics of limited firms might be an important direction for our future study.

\section*{Acknowledgments}

The authors wish to thank Dr.\ Bo Huang for the beneficial discussions and are grateful to the anonymous referees for their helpful comments.

This work has been supported by Philosophy and Social Science Foundation of Guangdong under Grant No.\ GD21CLJ01, Major Research and Cultivation Project of Dongguan City University under Grant Nos.\ 2021YZDYB04Z and 2022YZD05R, National Natural Science Foundation of China under Grant No.\ 11601023, and Beijing Natural Science Foundation under Grant No.\ 1212005.

\section*{Declaration of competing interest}

{The authors declare no conflict of interest.}

\section*{Appendix}

\begin{align*}

\begin{autobreak}
SP=
(972K^8
+19440K^7
+127575K^6
+162000K^5
-1552500K^4
-6412500K^3
-5062500K^2
+23437500K
+67187500
)(
8503056K^{12}
+191318760K^{11}
+1523464200K^{10}
+3754532250K^9
-14134854375K^8
-101982543750K^7
-146939062500K^6
+399469218750K^5
+1522072265625K^4
+261457031250K^3
-4576816406250K^2
-1938867187500K
+13981445312500),
\end{autobreak}\\

\begin{autobreak}
C_4(K,d)=
K^{8} d^{8}
+(
-126K^{8}
-210K^{7}) d^{7}
+(6660K^{8}
+21300K^{7}
+17800K^{6}) d^{6}
+(
-192024K^{8}
-874800K^{7}
-1382400K^{6}
-731000K^{5}) d^{5}
+(3285360K^{8}
+18688320K^{7}
+41115600K^{6}
+39438000K^{5}
+13350000K^{4}) d^{4}
+(
-33957792K^{8}
-221940000K^{7}
-588016800K^{6}
-728172000K^{5}
-379740000K^{4}
-45500000K^{3}) d^{3}
+(206172864K^{8}
+1453101120K^{7}
+4191652800K^{6}
+5433912000K^{5}
+2183760000K^{4}
-1105200000K^{3}
-478000000K^{2}) d^{2}
+(
-672686208K^{8}
-4870886400K^{7}
-14246409600K^{6}
-16185744000K^{5}
+2054160000K^{4}
+13262400000K^{3}
-7632000000K^{2}
-11520000000K) d
+906992640K^{8}
+6500113920K^{7}
+18223833600K^{6}
+13351392000K^{5}
-25284960000K^{4}
-27302400000K^{3}
+65376000000K^{2}
+30720000000K
-102400000000.
\end{autobreak}

\end{align*}

\bibliographystyle{abbrv}
\bibliography{monopoly.bib}

\end{CJK}
\end{document}